\documentclass[12pt,a4paper]{article}
\pdfoutput=1
\usepackage{jcappub}

\hypersetup{colorlinks,bookmarksopen,bookmarksnumbered,citecolor=blue,
linkcolor=black,pdfstartview=FitH,urlcolor=blue}

\usepackage{graphicx}
\usepackage{amsmath}
\usepackage{amssymb}
\usepackage{braket}
\usepackage{bm}

\def\bea{\begin{eqnarray}}
\def\eea{\end{eqnarray}}
\def\ba{\begin{eqnarray}}
\def\ea{\end{eqnarray}}
\def\be{\begin{equation}}
\def\ee{\end{equation}}
\def\beq{\begin{equation}}
\def\eeq{\end{equation}}

\newcommand{\lsim}{\mathrel{\rlap{\lower4pt\hbox{\hskip1pt$\sim$}}
    \raise1pt\hbox{$<$}}}         
\newcommand{\gsim}{\mathrel{\rlap{\lower4pt\hbox{\hskip1pt$\sim$}}
    \raise1pt\hbox{$>$}}}         

\newcommand{\leftrightarrowraised}{\mathrel{\rlap{\lower-0pt\hbox{\hskip1pt$\partial$}}
    \raise6 pt\hbox{$\leftrightarrow$}}}

\newcount\hour \newcount\minute
\hour=\time \divide \hour by 60
\minute=\time
\subheader{ADP-16-48/T1004}

\title{Self-interacting inelastic dark matter: A viable solution to the small scale structure problems}

\def\kth{Department of physics,
School of Engineering Sciences, KTH Royal Institute of Technology,
AlbaNova University Center, 106 91 Stockholm, Sweden}

\def\coepp{
ARC Center of Excellence for Particle Physics at the Terascale (CoEPP), University of Adelaide, Adelaide, SA 5005, Australia}

\author[a]{\textbf{Mattias Blennow,}\vspace*{0mm}}
\author[a]{\textbf{Stefan Clementz,}\vspace*{0mm}}
\author[a,\,b]{\textbf{Juan Herrero-Garcia}\vspace*{0mm}}
\affiliation[a]{\kth}
\affiliation[b]{\coepp}
  
\abstract{Self-interacting dark matter has been proposed as a solution to the small-scale structure problems, such as the observed flat cores in dwarf and low surface brightness galaxies. If scattering takes place through light mediators, the scattering cross section relevant to solve these problems may fall into the non-perturbative regime leading to a non-trivial velocity dependence, which allows compatibility with limits stemming from cluster-size objects. However, these models are strongly constrained by different observations, in particular from the requirements that the decay of the light mediator is sufficiently rapid (before Big Bang Nucleosynthesis) and from direct detection. A natural solution to reconcile both requirements are inelastic endothermic interactions, such that scatterings in direct detection experiments are suppressed or even kinematically forbidden if the mass splitting between the two-states is sufficiently large. Using an exact solution when numerically solving the Schr\"odinger equation, we study such scenarios and find regions in the parameter space of dark matter and mediator masses, and the mass splitting of the states, where the small scale structure problems can be solved, the dark matter has the correct relic abundance and direct detection limits can be evaded.}


\begin{document}
\maketitle

\section{Introduction}
The standard view that dark matter (DM) is collisionless and cold is currently very successful in its description of the universe. However, while N-body simulations of collisionless cold dark matter (CDM) predict cuspy radial profiles in which the density diverges as $r^{-\gamma}$, where $\gamma \gtrsim 1$~\cite{Navarro:1995iw,Moore:1997sg,Klypin:2000hk,Diemand:2005wv,Springel:2008cc}, observations of the DM halos of dwarf and low surface brightness galaxies as well as galaxy clusters point to flat cores~\cite{Moore:1994yx,Flores:1994gz,deBlok:2001hbg,deBlok:2002vgq,Swaters:2002rx,Simon:2004sr,Spekkens:2005ik,KuziodeNaray:2007qi,Spano:2007nt,deBlok:2008wp,Newman:2009qm,Donato:2009ab,Oh:2010ea,Walker:2011zu,Newman:2012nw,Adams:2014bda}.

One possible solution is self-interacting dark matter (SIDM)~\cite{Spergel:1999mh, Firmani:2000ce}\footnote{Another interesting solution to the above problems may be ultra-light scalars (with mass $\sim 10^{-22}$~eV) which can produce extended cores in dwarf galaxies, see e.g. refs.~\cite{Hu:2000ke,Goodman:2000tg,Marsh:2013ywa,Schive:2014dra,Schive:2014hza,Marsh:2015wka}.}. The main idea behind SIDM is that the DM particles that are trapped in the center of DM halos will gain kinetic energy by scattering with high velocity DM particles that are falling into the gravitational potential. These collisions will increase the temperature of the central parts, resulting in a less dense core. Numerical simulations studying SIDM have shown that a self-scattering cross section within roughly an order of magnitude of $\sigma_{\chi \chi}/m \sim 1 \,{\rm cm^2 / g}$ is capable of lowering the density in small scale DM halos~\cite{Dave:2000ar,Zavala:2012us,Rocha:2012jg,Elbert:2014bma,Vogelsberger:2015gpr,Vogelsberger:2012ku,Fry:2015rta}. However, the self-scattering cross section is constrained by observations of colliding galaxy clusters, the strongest constraint being~$\sigma_{\chi \chi}/m \lesssim 0.5$~cm$^2$/g~\cite{Harvey:2015hha,Randall:2007ph}. Other slightly weaker bounds come from studying the shapes of DM cluster halos~\cite{Peter:2012jh}, which can be fulfilled by SIDM models with cross sections that decrease with velocity~\cite{Loeb:2010gj,Tulin:2013teo, Kaplinghat:2015aga}.

Cross sections that decrease with velocity and thus evade the bounds arises naturally in models with light mediators. These are subject to different phenomenological constraints. In particular, for massive dark photons there are very strong bounds from supernovae cooling for masses in the lower MeV range~\cite{Dent:2012mx,Kazanas:2014mca,Rrapaj:2015wgs,Chang:2016ntp,Hardy:2016kme}. These scenarios are also constrained by Big Bang Nucleosynthesis (BBN) limits, which demand that the mediators decay sufficiently early~\cite{Kaplinghat:2013yxa}. However, this requirement is typically incompatible with the strong limits from direct detection~\cite{Batell:2009vb,Kaplinghat:2013yxa,DelNobile:2015uua}. Another implication of these models is the Sommerfeld enhancement of the annihilation cross section, which significantly constrains DM models with s-wave annihilations due to their energy injection in the plasma at the time of recombination~\cite{Bringmann:2016din}. The latter constraints can be evaded with DM that has p-wave annihilations.

A natural way to avoid direct detection (DD) bounds are inelastic DM models, in which there are at least two different states with a small mass difference~\cite{TuckerSmith:2001hy,TuckerSmith:2004jv,Chang:2008gd,Graham:2010ca,An:2011uq,Schwetz:2011xm,McCullough:2013jma,Fox:2013pia,Bozorgnia:2013hsa,Frandsen:2014ima,Chen:2014tka,Blennow:2015hzp}. In principle, both endothermic (up-scattering), and exothermic (down-scattering) interactions could take place. However, as we will investigate in sec.~\ref{abundance}, in the case of large self-interactions as required, it is in general the lowest mass state which gives the dominant contribution to the relic abundance.

Therefore, if the splitting is sufficiently large, endothermic scatterings in direct detection detectors are kinematically forbidden. Even if kinematically allowed, inelastic DM weakens the constraints (for instance from xenon experiments like PANDA~\cite{Tan:2016zwf} and LUX~\cite{Akerib:2015rjg}, and from germanium ones like CDMSlite~\cite{Agnese:2015nto}) by probing only the high velocity tail of the galactic DM velocity distribution, as only DM particles above a larger velocity (compared to the elastic scattering case) can produce detectable recoils. The phenomenology of inelastic DM has been explored in various astrophysical contexts and recently also in regard to self-scattering interactions as a solution to the small scale structure puzzles. The s-wave case with a low mass mediator is treated in ref.~\cite{Schutz:2014nka}, a very large mass splitting is considered in ref.~\cite{Zhang:2016dck} and a similar approach with atomic DM which is inelastic due to hyperfine splittings has been carried out in ref.~\cite{Boddy:2016bbu}. In this work we aim to study the DM self-scatterings by solving the Schr\"odinger equation with the same type of non-diagonal Yukawa-potential considered for instance in refs.~\cite{Schutz:2014nka,Zhang:2016dck}. Unlike previous studies, instead of resorting to approximations or considering a single partial wave to avoid the instabilities that occur in the numerical solutions, we will reformulate the Schr\"odinger equation in a way that allows for a much more stable numerical solution.

There are also other interesting effects of inelastic DM that may alleviate or even solve these puzzles. As noted in ref.~\cite{Loeb:2010gj}, de-excitations of two excited state particles may occur. If the mass splitting is large, the released kinetic energy may leave the two outgoing DM particles with velocities much greater than the escape velocity of the object they were originally bound to. These particles may then either re-scatter and become gravitationally bound, resulting in a hotter halo, or escape, resulting in halo evaporation. The excited state may also be unstable and decay into the lower state after a halo has formed, emit some light particle and pick up some recoil in the process, with the net result being essentially the same as in down-scattering~\cite{SanchezSalcedo:2003pb,Abdelqader:2008wa,Peter:2010au,Peter:2010jy,Peter:2010sz,Bell:2010qt}. The importance of these processes will also depend crucially on the abundance of the excited state.

The paper is structured as follows. In sec.~\ref{framework} we describe the phenomenological framework for inelastic DM with large self-interactions that we consider, and in sec.~\ref{sec:pheno} we summarize the most relevant phenomenological constraints. In sec.~\ref{results}, we compute the elastic and inelastic self-interacting cross sections by numerically solving the Schr\"odinger equation, focusing on the regions of parameter space that are capable of solving the small scale structure problems. We give further details of the procedure employed in appendices~\ref{sec:details} and \ref{sec:num_sol}. In sec.~\ref{abundance} we study the relative relic abundance of the heavier state. Finally, we discuss the results and give our conclusions in sec.~\ref{conc}.

\section{Inelastic DM framework} \label{framework}

There are many models in which DM is inelastic, see e.g. ref.~\cite{Cui:2009xq}. Generally, these type of models consist of at least two different DM states $\chi_i$ and $\chi_j$ whose masses satisfy $m_j - m_i = \delta \ll m_j , m_i$. The inelasticity stems from the fact that the interaction vertices takes one state into another. We study models of DM with masses within the range $[1, 10^3]$~GeV, and mass splittings in the range $10\, {\rm keV} \leq |\delta| \leq 1\, {\rm MeV}$.

We assume in all generality a mediator with mass $m_{A^\prime}$, in the mass range $[0.1,\,10^3]$ MeV, with a dark fine structure constant $\alpha_\chi= g_\chi^2/(4\pi)$. The mixing with the SM particles, can come either from mixing with the Higgs for scalar mediators, or via kinetic mixing with photon or mass mixing with the Z-boson for spin-1 mediators, see for instance ref.~\cite{Kaplinghat:2013yxa}.

When the force amongst DM particles are carried by a light mediator, the force will be long-ranged, and if the dark coupling constant is large, the perturbative calculation of the cross section will break down. In this region, the scattering cross section can be found by solving the Schr\"odinger equation for the DM wave-function $\Psi(\vec{x})$
\begin{equation}
\left[ -\frac{\nabla^2}{m_\chi} + V(\vec{x}) - \frac{k^2}{m_\chi} \right]\Psi(\vec{x}) = 0\,,
\end{equation}
where $m_\chi$ is the DM mass and $k = m_\chi v/2$ the momentum of either incoming particle in the center of mass frame, and the relevant potential $V(\vec{x})$ is a $2 \times 2$ matrix given by
\begin{equation} \label{V_pot}
V(r) = \left(
  \begin{array}{ c c }
     0 & -\frac{\alpha_\chi e^{-m_{A'} r}}{r} \\
     -\frac{\alpha_\chi e^{-m_{A'} r}}{r} & 2 \delta
  \end{array} \right)\,,
\end{equation}
which is shown to arise for example in models with two Majorana fermions and a massive dark photon~\cite{ArkaniHamed:2008qn,Zhang:2016dck}. In the simple inelastic case considered here, $\Psi$ is a $2 \times 1$ vector which describes the $\chi \chi$ content (first component) and $\chi^* \chi^*$ content (second component). While we focus on $\chi \chi$ and $\chi^* \chi^*$ scattering, we note that for the dark photon scenario, there will also be a $2 \times 2$ Schr\"odinger equation that describes elastic scattering $\chi \chi^*$ for which one can, without any problems due to the mass splitting, apply the general procedure outlined in ref.~\cite{Tulin:2013teo}. In other models, there may even be other possible scattering channels such as $\chi \chi^* \rightarrow \chi \chi$, see e.g., refs.~\cite{Finkbeiner:2008gw,Finkbeiner:2009mi}.

Using the method of separation of variables and expanding the Schr\"odinger equation in partial waves, one finds that each partial wave $l$ will satisfy the equation
\begin{equation} \label{partial_schr_eqn}
\left[ \frac{1}{r^2}\partial_r \left( r^2 \partial_r \right) - \frac{l(l+1)}{r^2} + k^2 \right] R_{l,i}(r) = m_\chi V_{ij}(r)R_{l,j}(r).
\end{equation}
Solving the system numerically is difficult in the parts of the parameter space where at least one state is kinematically inaccessible, which happens in our case when $\chi \chi \rightarrow \chi^* \chi^*$ scattering is forbidden. The numerical instability is due to exponential growth of the wave-function at large $r$ of any channel for which $k_i^2 < 0$. The stability is greatly enhanced due to the series of substitutions presented in appendices~\ref{sec:details} and \ref{sec:num_sol}, which are also applicable to models with more channels and an arbitrary potential.

An interesting note is that elastic self-interaction is not exclusively taking place for $\chi \chi^*$ scattering but both $\chi \chi \rightarrow \chi \chi$ and $\chi^* \chi^* \rightarrow \chi^* \chi^*$ are allowed. In other words, any type of scattering may take place except up-scattering when it is kinematically forbidden.

If the relic abundances of $\chi$ and $\chi^*$ occur through the standard freeze-out scenario, there will be no distinction between the two states at the temperature at which this occurs indicating that both states are equally populated. Imposing that the annihilation cross section $\sigma_{\rm ann}$ is equal to the thermal freeze-out value for weak scale DM (assuming that they make up the whole DM abundance), $\alpha_\chi$ is fixed in terms of the DM mass $m_\chi$. Neglecting the Sommerfeld enhancement, which at the high velocity of DM particles at freeze-out is at most an $\mathcal{O}(1)$ effect~\cite{Slatyer:2009vg} in the part of the parameter space that we consider, and the $\mathcal{O}(1)$ factor depending in the nature of the final state mediators, one gets for s-wave annihilation: 
\be \label{thermal}
\sigma_{\rm ann}\simeq \frac{\pi\,\alpha_\chi^2}{m^2_{\chi}} \sim 10^{-9}\,\text{GeV}^{-1}\,.
\ee
Note that in our numerical study we will use the value of ref.~\cite{Batell:2009vb},
\begin{equation} \label{dark_alpha}
\alpha_\chi = 0.01 \left( \frac{m_\chi}{270\,{\rm GeV}} \right)\,,
\end{equation}
unless otherwise stated. If annihilation takes place through a p-wave process, the annihilation cross section at freeze-out will be velocity suppressed which slightly increases $\alpha_\chi$ in terms of the DM mass, $\alpha_\chi \sim 0.01 (m_\chi / 100$~GeV)~\cite{Tulin:2013teo}.

\subsection{Phenomenological constraints} \label{sec:pheno}
There are some constraints one should take into consideration in models with light mediators, in the mass range $0.1-1000$ MeV. If the mediator is stable, it may easily overclose to universe. If it is unstable, it should decay before the onset of Big Bang Nucleosynthesis or it may lead to inconsistent primordial element abundances. Therefore, if its mass is below the MeV, it should decay into neutrinos or other dark sector particles. On the other hand, direct detection limits strongly bound the mixing between the dark mediators and the SM photon, Z or Higgs. These opposite requirements strongly bound the allowed parameter space models, basically excluding the scalar mediator case~\cite{Bringmann:2016din}.

\begin{figure}
	\centering
	\includegraphics[width=0.4\textwidth]{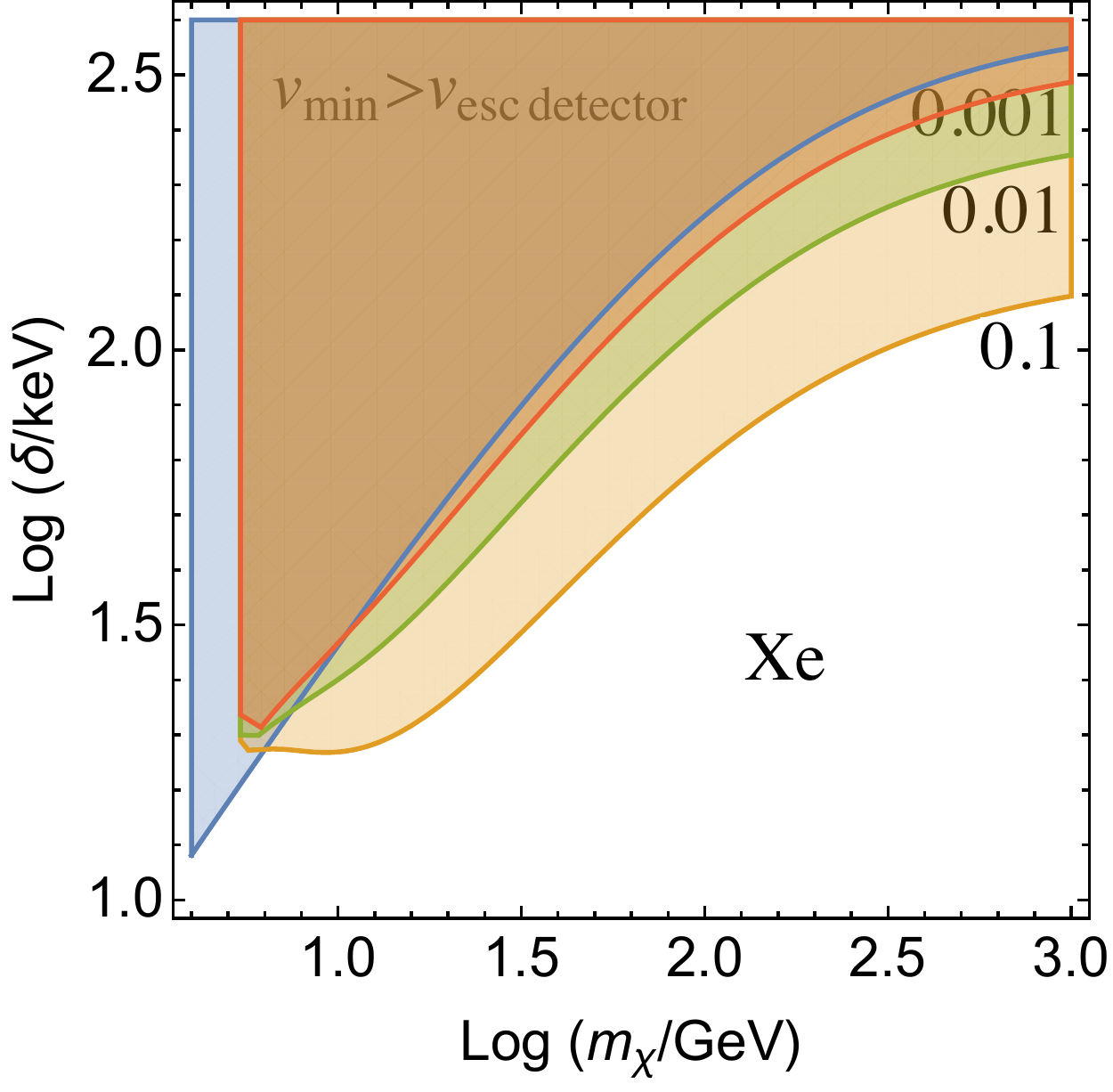}\qquad
	\includegraphics[width=0.4\textwidth]{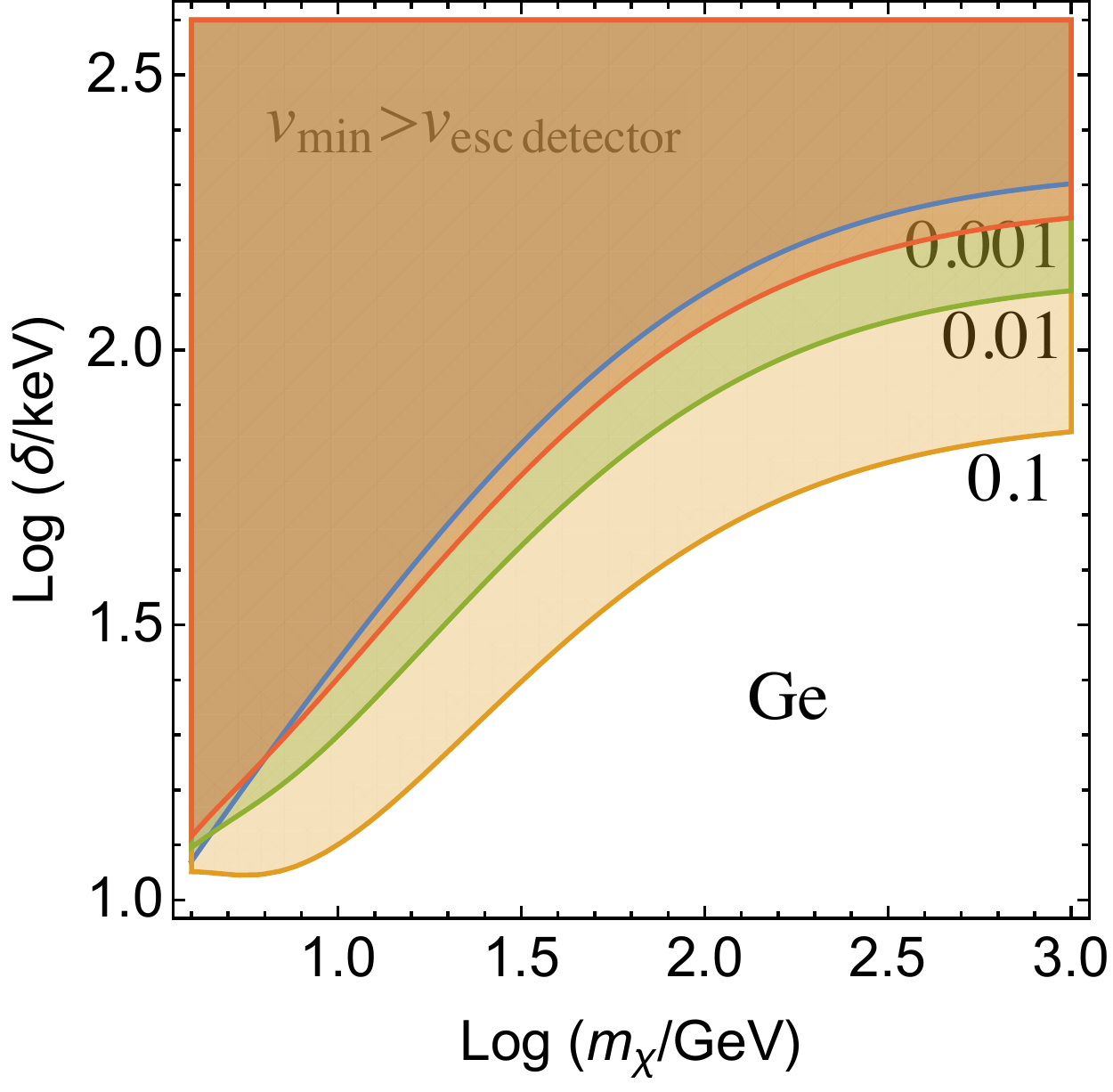}
		\caption{Contours (in blue) of $v_{\rm min}=v_{\rm esc\,detector}=750\,{\rm km/s}$ (in the detector rest-frame) in the plane Log($\delta/{\rm keV}$) - Log($m_\chi/{\rm GeV}$), shown for xenon (left) and germanium (right). The shaded region above it is kinematically forbidden. We also show contours of the conservative lower bounds on the suppressions of the inelastic rate with respect to the elastic one ($10,\,1,\,0.1\,\%$ from bottom to top), i.e., for every $\delta$ and $m_\chi$ the suppression will be larger than the quoted values.} \label{inel_plots}
\end{figure}

Therefore, inelastic scatterings of DM with a sufficiently large splitting are a natural way to reconcile sufficiently large mediator decay widths with absence of DD signals (see also ref.~\cite{Zhang:2016dck}). The minimum velocity required for a DM particle to give rise to endothermic scattering in a DD experiment is related to the mass splitting and the DM mass by $v_{\rm min} = \sqrt{2 \delta / \mu_{\chi A}}$ where $\mu_{\chi A}$ is the reduced mass of the DM-nucleus system. Reducing the DM mass or increasing the mass splitting both increases the required minimum velocity. We plot in fig.~\ref{inel_plots} the splitting $\delta$ versus the DM mass $m_\chi$ for xenon (left) and germanium (right) experiments such that $v_{\rm min} >v_{\rm esc\, detector} \simeq 750\,{\rm km/s}$, where $v_{\rm esc\, detector}$ is the escape velocity in the detector rest-frame, and therefore scatterings are kinematically forbidden. We also show contours where the inelastic rate \footnote{The contours are the integration of a Maxwell-Boltzmann distribution, with mean velocity $v_{s}=220\,{\rm km/s}$, from the minimum velocity of each nucleus. The rates will be modified for spin-independent and spin-dependent due to the form factors, which are not included here. For the elastic case we use threshold energies of $2\,(3)$ keV for germanium (xenon).} is at least $0.1,\,1,\,10\,\%$ of the elastic one and thus the upper limits on the cross section would be at least $1,\,2,\,3$ orders of magnitude weaker (and therefore the width of the mediator could be larger by the same amount).  One can see that scatterings are kinematically forbidden in xenon for $m_\chi =10\,(100)$~GeV if $\delta \gtrsim (30)\,180$~keV, and in germanium if $\delta \gtrsim 25\,(125)$~keV. In both cases, for any DM mass, significant suppressions occur, i.e., larger than $10\,\%$,  for $\delta\gtrsim60\,(100)$  keV for xenon (germanium).

In scenarios of large self-scattering cross sections due to long-rang interactions, Sommerfeld enhancements can give rise to large annihilations. For the potential we consider, ref.~\cite{Slatyer:2009vg} finds that the annihilation cross section can be significantly enhanced at low velocities. Therefore, DM annihilation in dwarf galaxies, or its imprint on the CMB, impose severe limits on the allowed parameter space of the models. In general, S-wave annihilations are very constrained~\cite{Tulin:2013teo, Bringmann:2016din}, while p-wave annihilations, dominant decays into neutrinos (mass mixing with the Z) or a different dark sector temperature are viable options~\cite{Tulin:2013teo, Bringmann:2016din}. Other possibilities include DM that is produced non-thermally, via freeze-in~\cite{Bernal:2015ova} or MeV scale DM that annihilates directly into SM particles through a heavier mediator~\cite{Chu:2016pew}.\footnote{Asymmetric DM is of course also a way out for the case of elastic scatterings. For inelastic DM, it is difficult to envision such a model.}

\section{Results} \label{results}
In this section, we will discuss the numerical results from solving the Schr\"odinger equation. 

Elastic self-scattering cross sections roughly in the range $0.5 < \sigma_{\chi \chi}/m_\chi < 5$~cm$^2$/g have been shown to positively affect the small scale structure objects \cite{Dave:2000ar,Zavala:2012us,Rocha:2012jg,Elbert:2014bma,Vogelsberger:2015gpr}. The simple picture of these studies is complicated by the fact that halos in inelastic DM models may consist of not only $\chi$ but also of $\chi^*$. With mass splittings of the size considered here, up-scattering is typically forbidden in the low velocity objects. If a halo consist of only the lower mass state, one can make a direct comparison of the cross sections at low velocities. The presence of a significant fraction of $\chi^*$ in a halo changes the picture for two reasons. Down-scattering can occur which will boost the two $\chi$ that are produced in the collision. If these escape, the $\chi^*$ population of the halo can evaporate. If they rescatter with other $\chi$, this will result in an overall heating of the halo.

To assess to what extent elastic scattering may affect halos, we use the viscosity cross section $\sigma_V$ which down weights scattering in both the forward and the backward directions,
\begin{equation}
\sigma_V = \int \frac{d\sigma}{d\Omega}\,{\rm sin}^2(\theta) \, d\Omega \, .
\end{equation}
This weight should be used since forward and backward scattering do not alter the halo structure. This is due to the particles travelling in the same direction before and after the collision do not alter the halo. \footnote{Another commonly used weighed cross section is the transfer cross section, which cancels out forward scattering but maximizes backwards scattering~\cite{Loeb:2010gj,Tulin:2013teo},  $\sigma_T = \int d\sigma/d\Omega\, \left( 1- {\rm cos}(\theta) \right) d\Omega$. As noted in ref.~\cite{Boddy:2016bbu}, the viscosity cross section is more appropriate than the transfer one since the momentum transfer in the latter is weighed maximally towards back scattering which, for reasons already mentioned, is not interesting.}

For inelastic scattering, the full cross section is calculated since energy-loss or gain will be significant in these collisions. These are thus expected to have an affect on the typical halo regardless of the angular dependence of the collision.

The procedure of calculating the cross section from the solution of the Schr\"odinger equation is theoretically straight-forward, although the numerical solutions to the differential equation are unstable in a large part of the parameter space considered here. This can be remedied by a series of substitutions outlined in appendix~\ref{sec:num_sol}. We find good agreement with previous results that consider the same Yukawa-type potential in the $l=0$ case~\cite{Schutz:2014nka} and in the case where an adiabatic approximation is employed~\cite{Zhang:2016dck}. We notice that it takes significantly longer to calculate self-scattering of particles initially in the excited state.

\begin{figure}
	\centering
	\includegraphics[width=0.48\textwidth]{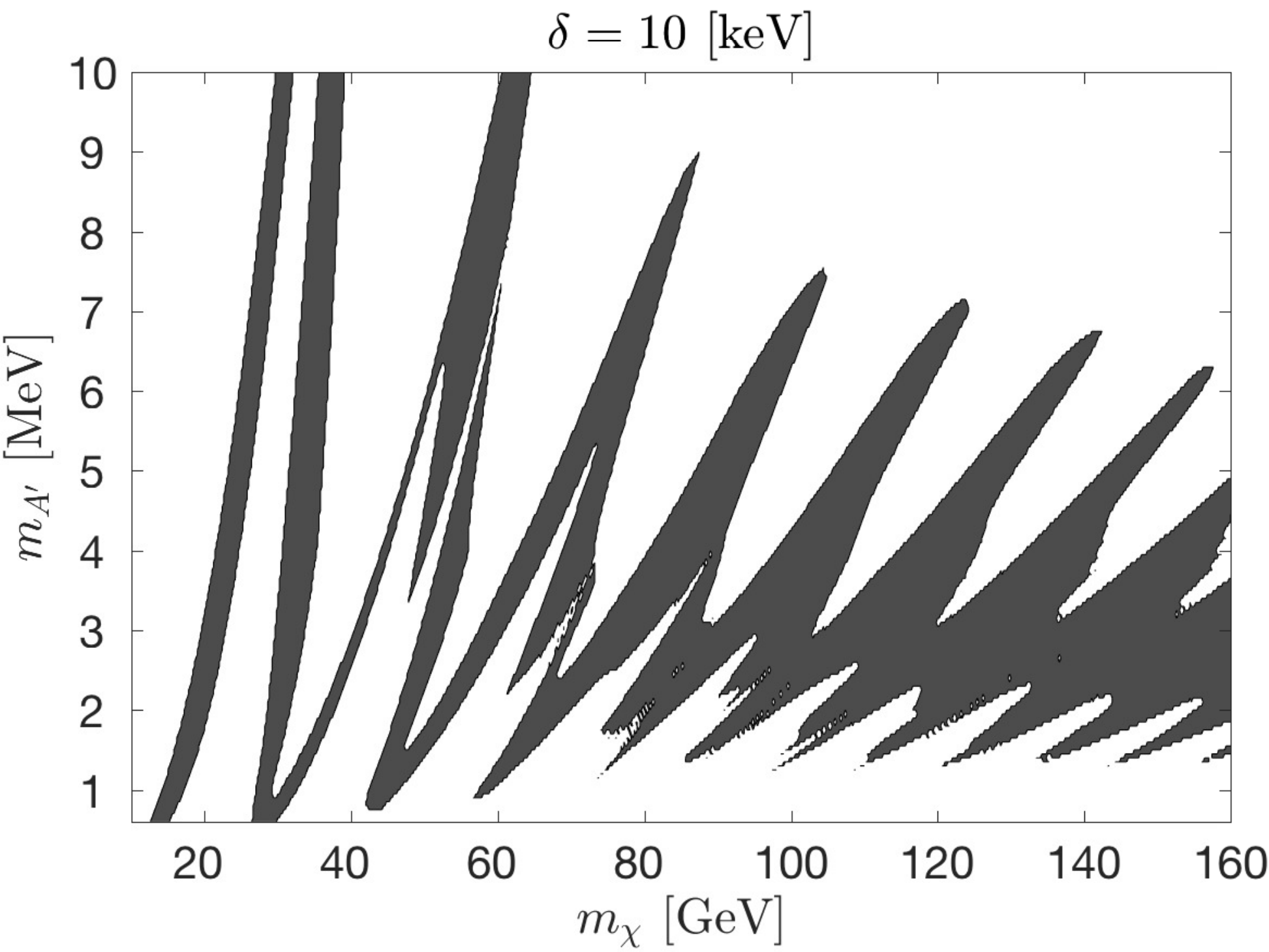}
	\includegraphics[width=0.48\textwidth]{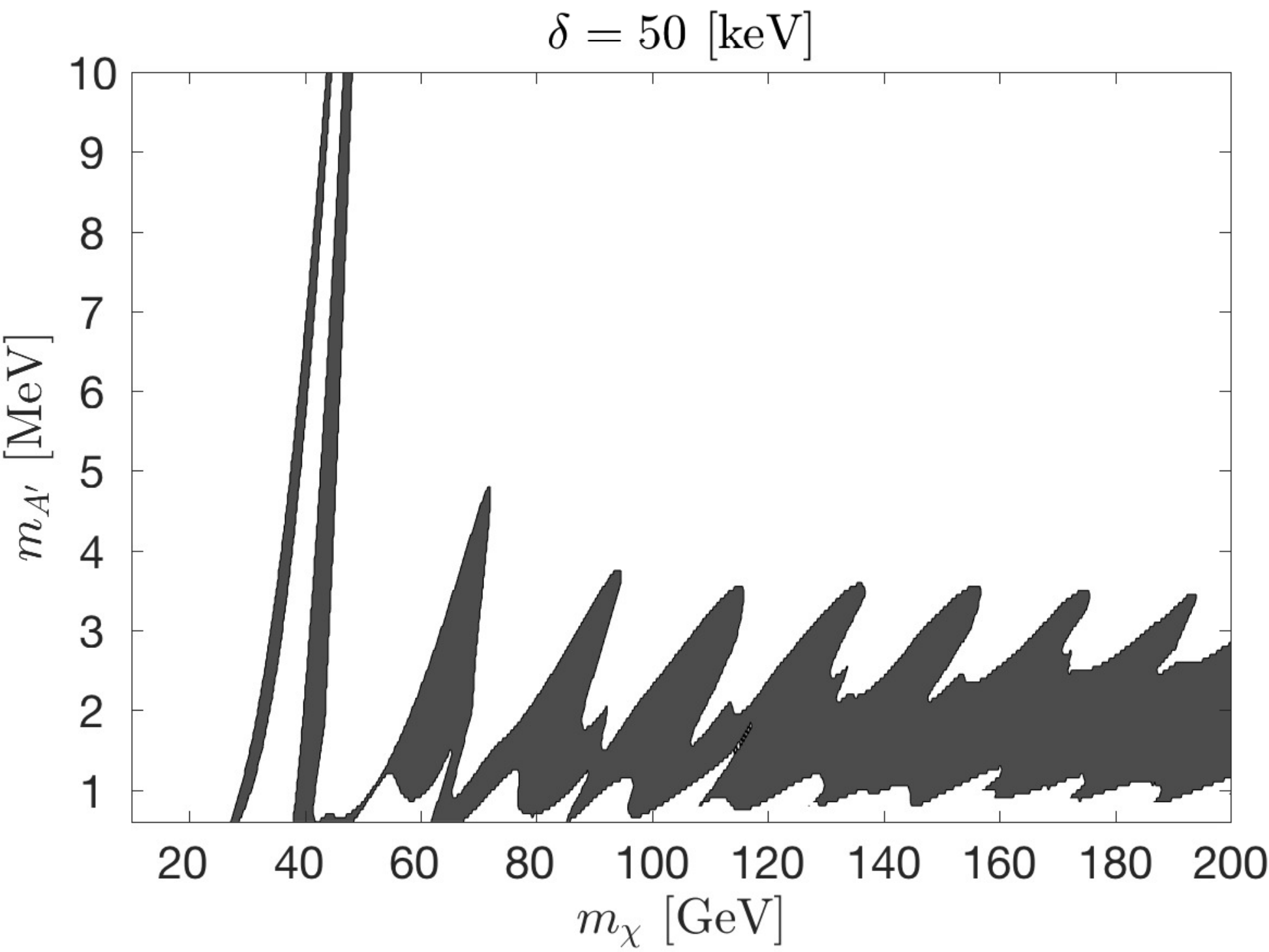}\\
	\vspace{3mm}
	\includegraphics[width=0.48\textwidth]{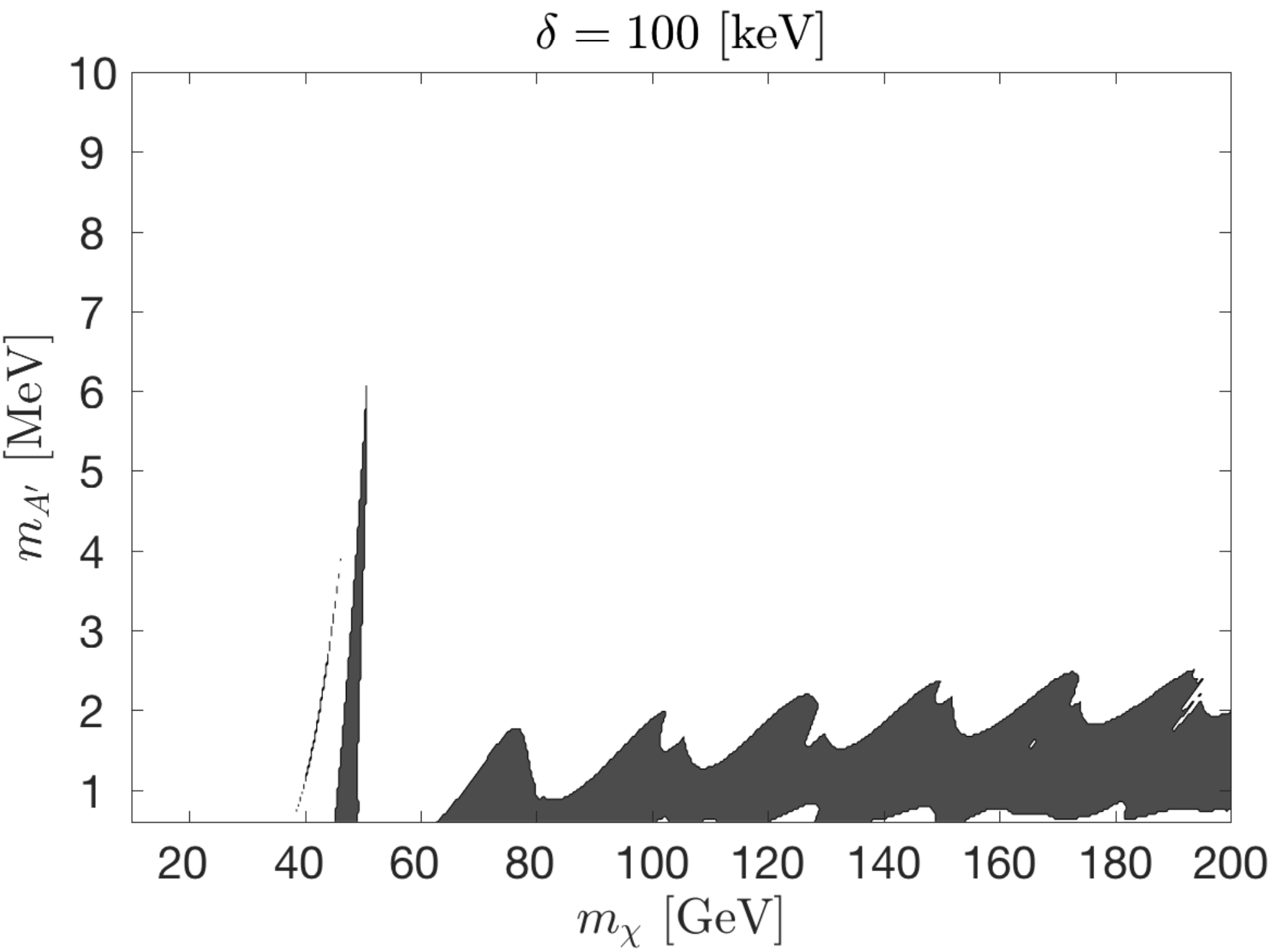}
	\includegraphics[width=0.48\textwidth]{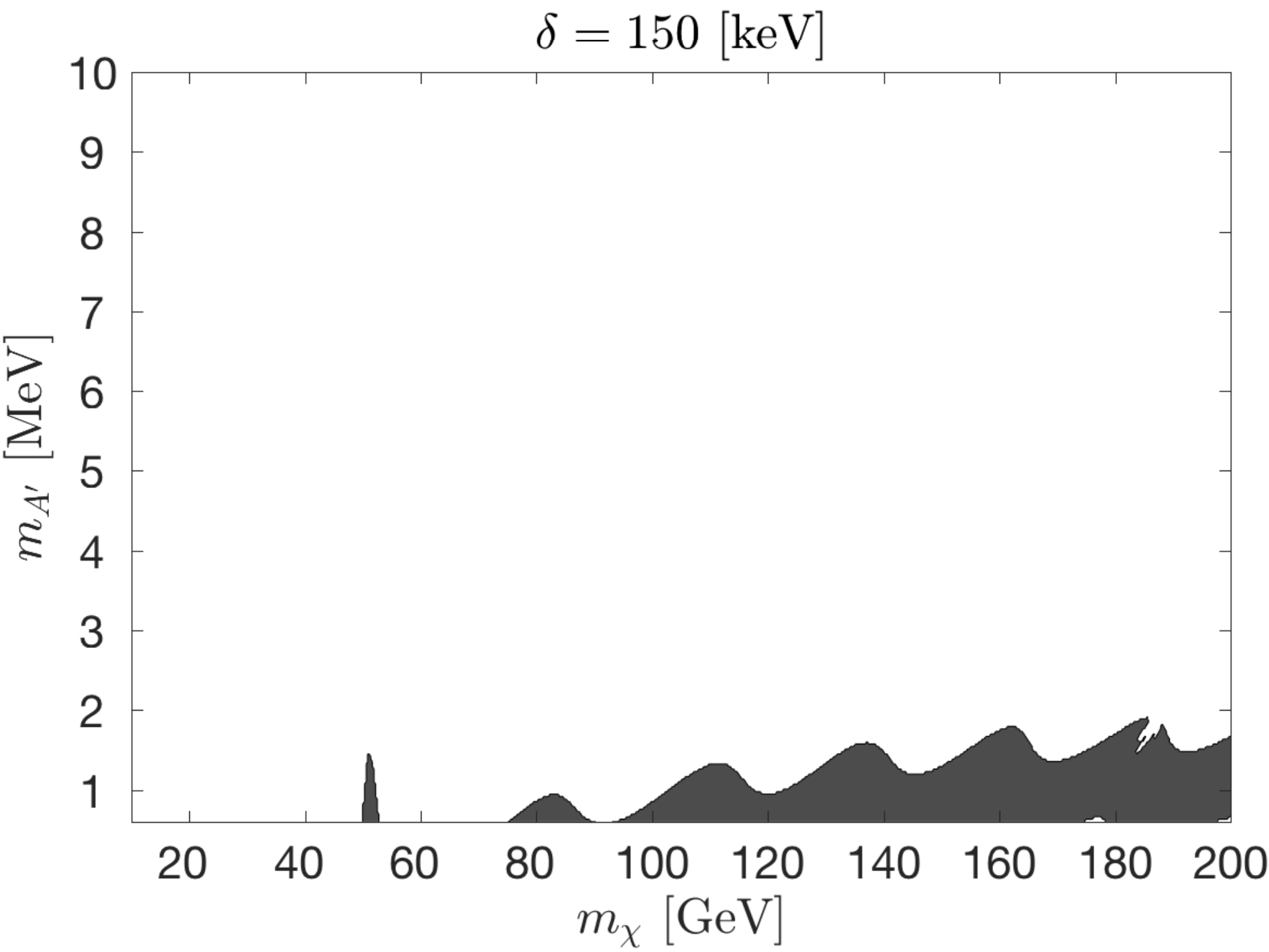}
	\caption{Greyed out is a region where $0.5<\sigma_{\chi \chi \rightarrow \chi \chi,V}<5$~cm$^2$/g is satisfied in a velocity interval relevant to dwarf and low surface brightness galaxies. The mass splitting is set to $\delta=10$~keV in the upper left, $\delta = 50$~keV in the upper right, $\delta = 100$~keV in the lower left and $\delta = 150$~keV in the bottom right plot.} \label{gridplots}
\end{figure}

In figure~\ref{gridplots}, we show a region in the $m_\chi-m_{A'}$~plane where the elastic viscosity cross section of the $\chi$'s, $\sigma_{\chi \chi \rightarrow \chi \chi , V}$ is consistent with cores in dwarf and low surface brightness galaxies. We take this to be where $0.5<\sigma_{\chi \chi \rightarrow \chi \chi,V}<5$~cm$^2$/g for velocities in the range $30 - 100$~km/s. The upper left plot shows the case where $\delta=10$~keV, the upper right plot $\delta=50$~keV, the bottom left plot $\delta = 100$~keV and the bottom right plot $\delta = 150$~keV. For lower DM masses, the cross section decreases with increasing $\delta$, while for larger DM masses it approaches a constant value. This is expected, since all cases tend to the same model when $\delta / m_\chi \rightarrow 0$.

\begin{figure}
	\centering
	\includegraphics[width=0.47\textwidth]{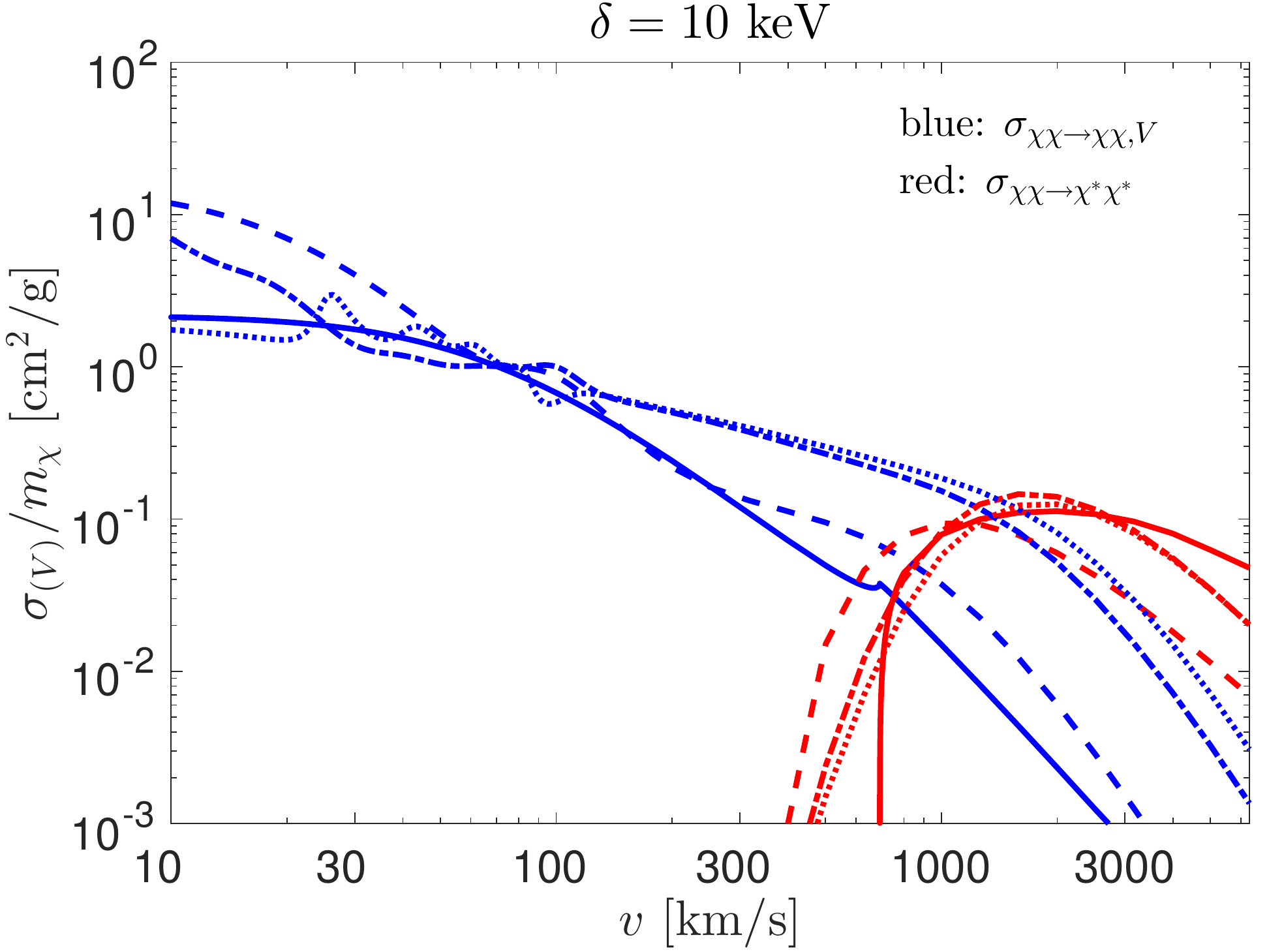}
	\hspace{1mm}
	\includegraphics[width=0.47\textwidth]{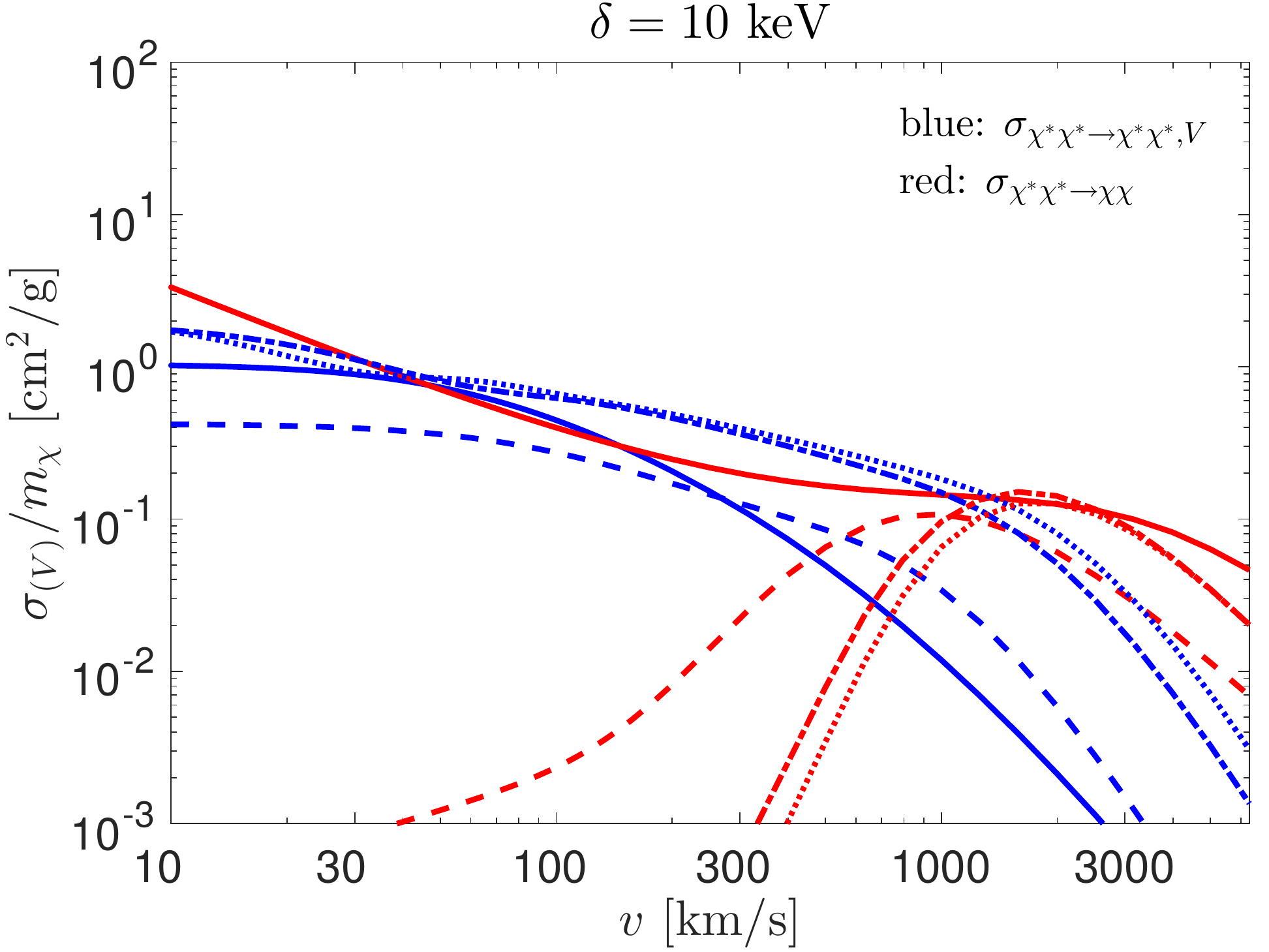}\\
	\vspace{3mm}
	\includegraphics[width=0.47\textwidth]{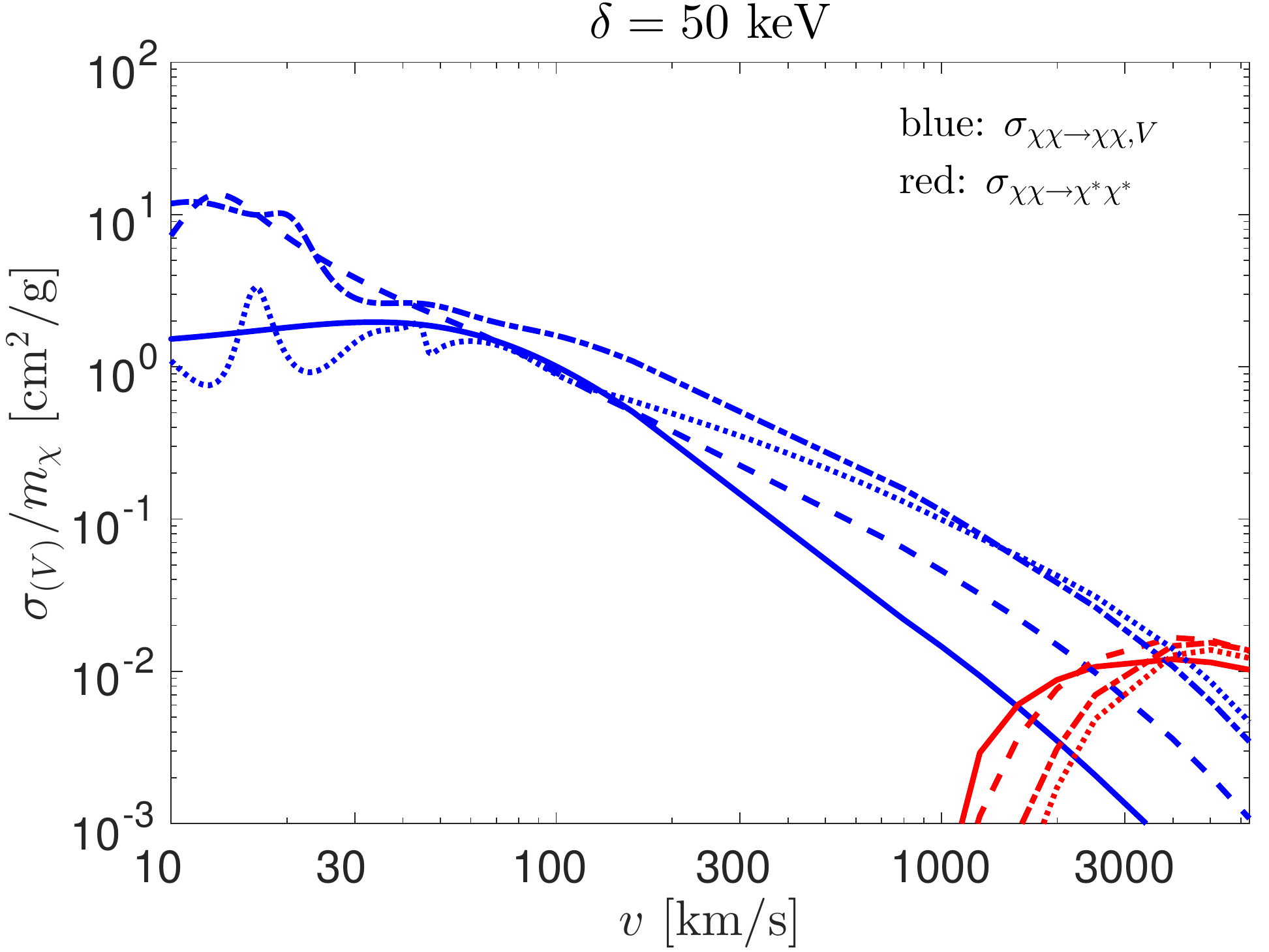}
	\hspace{1mm}
	\includegraphics[width=0.47\textwidth]{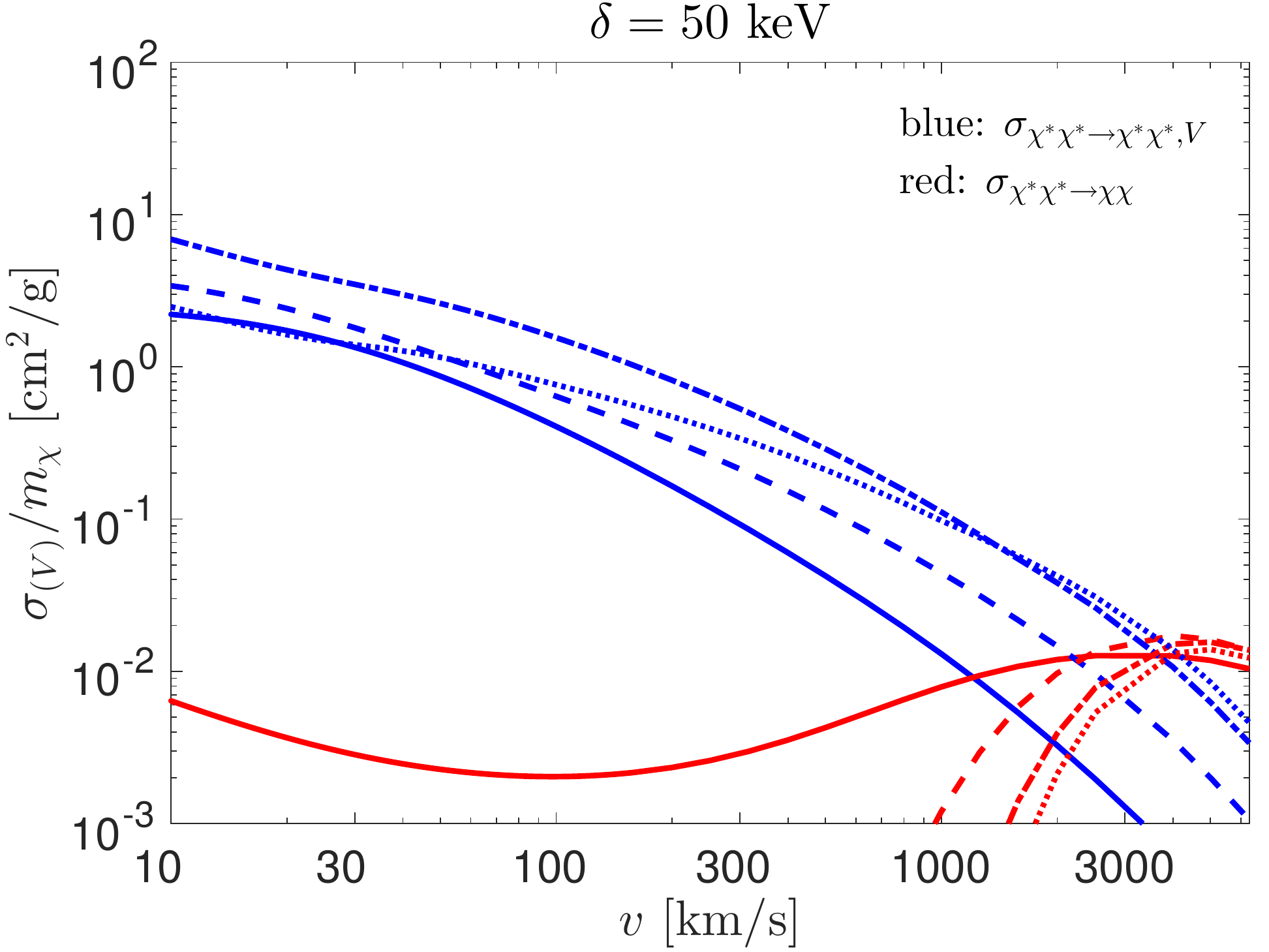}\\
	\vspace{3mm}
	\includegraphics[width=0.47\textwidth]{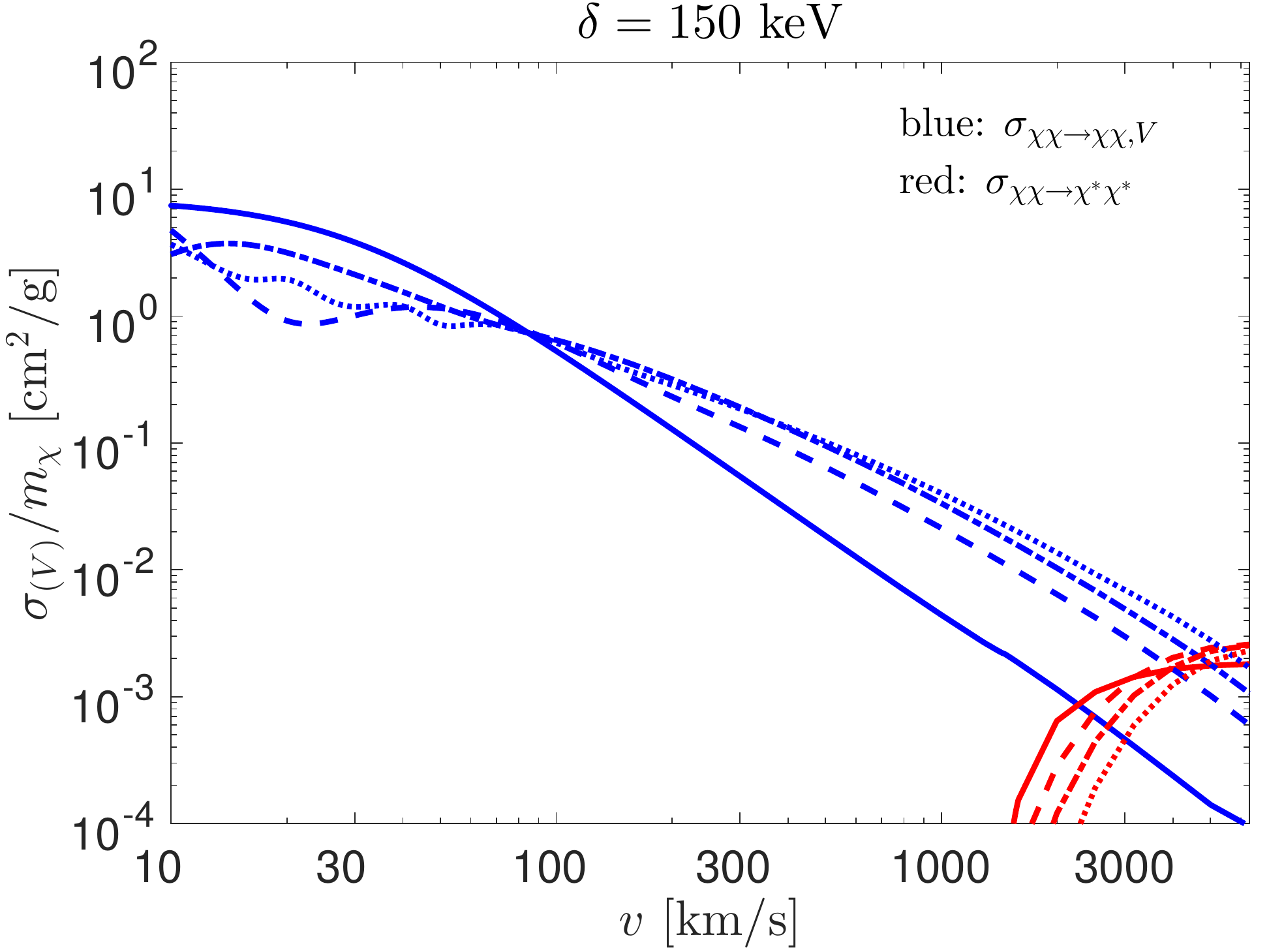}
	\hspace{1mm}
	\includegraphics[width=0.47\textwidth]{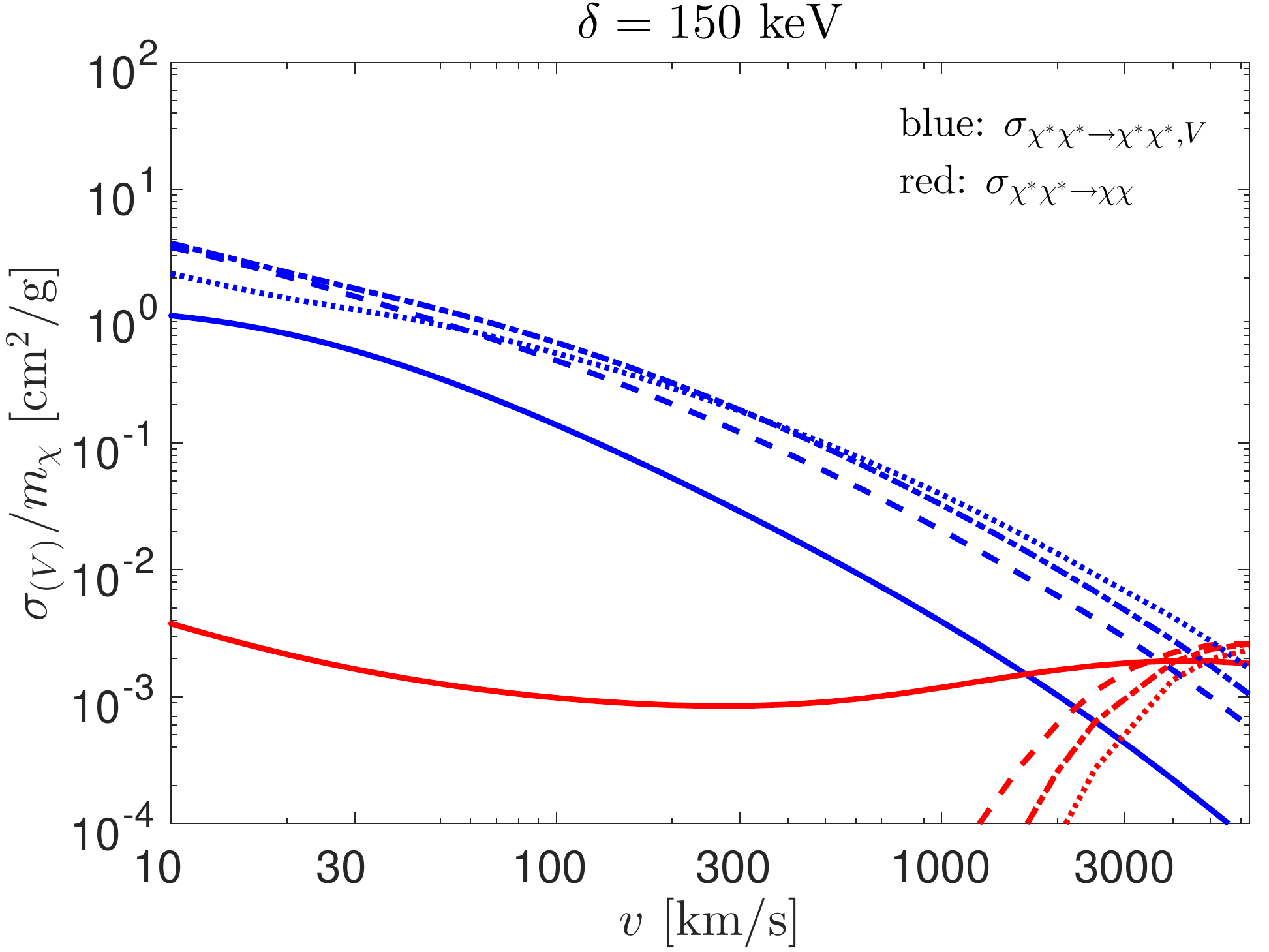}\\
	\caption{Elastic and inelastic self-scattering cross sections versus DM velocity, shown for various choices of $m_\chi$ and $m_{A'}$ for which the elastic interactions $\chi \chi$ and $\chi^* \chi^*$ can solve the small scale structure problems. The choice of $\delta$ are $10\, (50)\, [150]$~keV in the first (second) [third] rows. The left plots shows $\chi\chi$ scattering, while the right plots shows $\chi^*\chi^*$ scattering. The parameters chosen for each line are listed in table~\ref{benchmark_points}.} \label{sigma_examples}
\end{figure}

Next, we compute the scattering cross section as a function of velocity for some allowed benchmark points of $m_\chi$ and $m_{A'}$. The benchmark points chosen are listed in table~\ref{benchmark_points} and the results can be seen in fig.~\ref{sigma_examples} for $\delta=10\,(50)\, [100]$~keV in the upper (middle) [bottom] rows. The left plot shows scattering between two $\chi$ and the right plot shows scattering between two $\chi^*$ particles. In all cases, one can observe that the elastic scattering cross sections of $\chi$'s and $\chi^*$'s are similar in size. One can also see how, the larger the mass splitting, the larger the velocity at which inelastic endothermic (double excitation) scatterings kick in. For splittings larger than the MeV, this occurs at velocities larger than those that are typical in cluster halos.
 This is expected for up-scattering (double excitation) since it is kinematically forbidden, but the plots on the right show that also down-scattering (double de-excitation) can be significantly reduced with respect to elastic scattering at low velocities. This feature is most pronounced in the cases where the DM mass is larger. Only for the benchmark point A of $\delta = 10$~keV does the down-scattering cross section rival the elastic scattering cross section. 

\begin{table}
\centering
    \begin{tabular}{ c c  c | c  c | c  c |}
        \cline{2-7}
       &   \multicolumn{2}{| c |}{$\delta = 10$ keV} & \multicolumn{2}{| c |}{$\delta = 50$ keV} & \multicolumn{2}{| c |}{$\delta = 150$ keV} \\ \cline{1-7}
         	\cline{1-7}  \multicolumn{1}{|c|}{Benchmark (line style)} & \multicolumn{1}{|c}{$m_\chi$ } & \multicolumn{1}{c|}{$m_{A'}$ } & \multicolumn{1}{|c}{$m_\chi$ } & \multicolumn{1}{c|}{$m_{A'}$ } & 			\multicolumn{1}{|c}{$m_\chi$ } & \multicolumn{1}{c|}{$m_{A'}$ }\\
        		\cline{1-7} \cline{1-7}  \multicolumn{1}{|c|}{A (solid)} & 15 & 1 & 40 & 1 & 51 & 1\\
		\cline{1-7} \multicolumn{1}{|c|}{B (dashed)} & 55 & 7 & 80 & 1.5 & 110 & 1\\
		\cline{1-7} \multicolumn{1}{|c|}{C (dash-dotted)} & 100 & 4 & 120 & 1 & 140 & 1 \\
		\cline{1-7} \multicolumn{1}{|c|}{D (dotted)} & 140 & 4 & 160 & 2 & 180 & 1.5\\
        \hline
    \end{tabular}
\caption{Benchmark points chosen for each $\delta$ to compute the cross sections in fig.~\ref{sigma_examples}. The line style refers to the lines in the figures. DM masses are listed in GeV and mediator masses in MeV.}
\label{benchmark_points}
\end{table}

\begin{figure}
	\centering
	\includegraphics[width=0.48\textwidth]{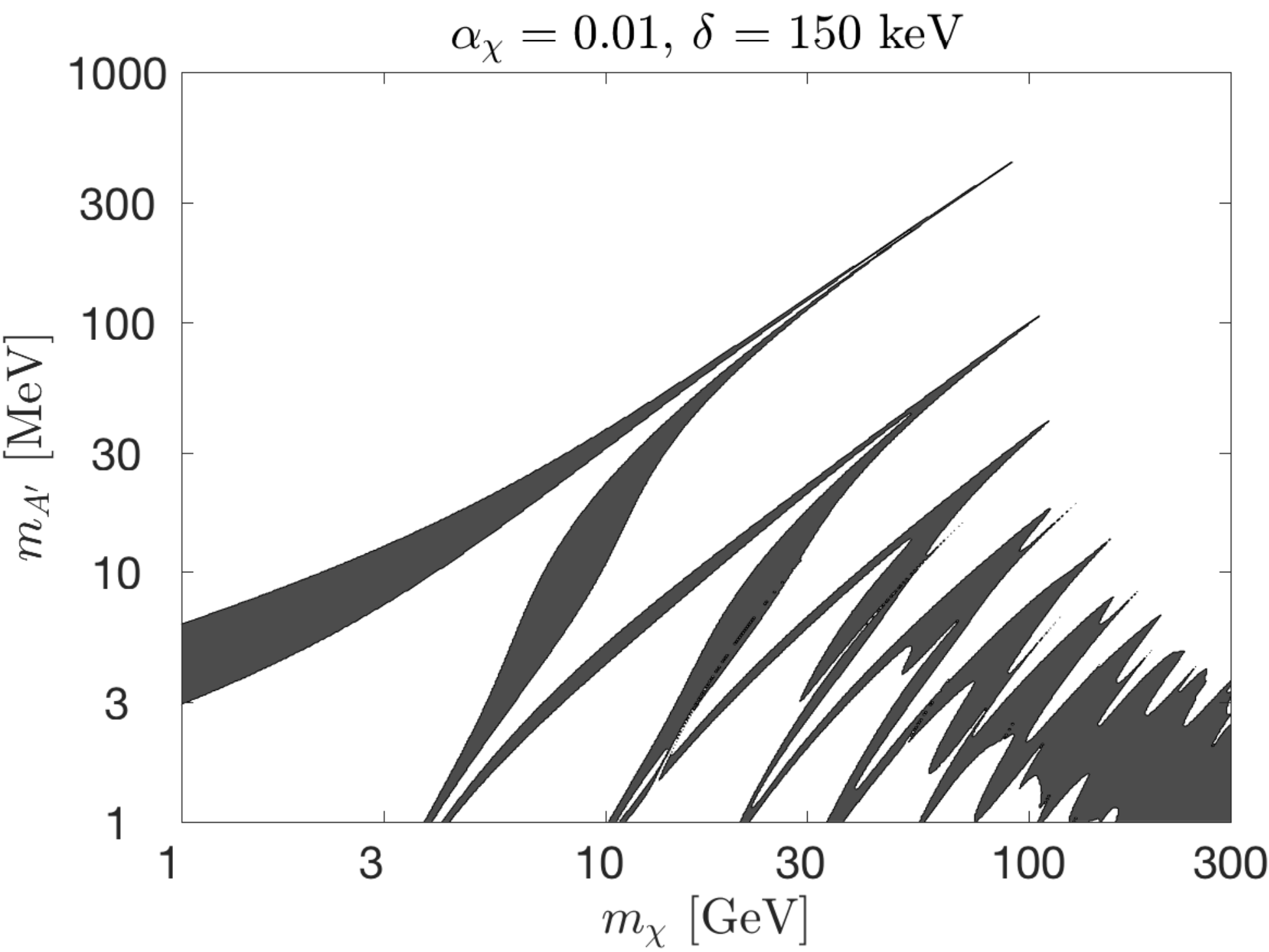}
	\includegraphics[width=0.48\textwidth]{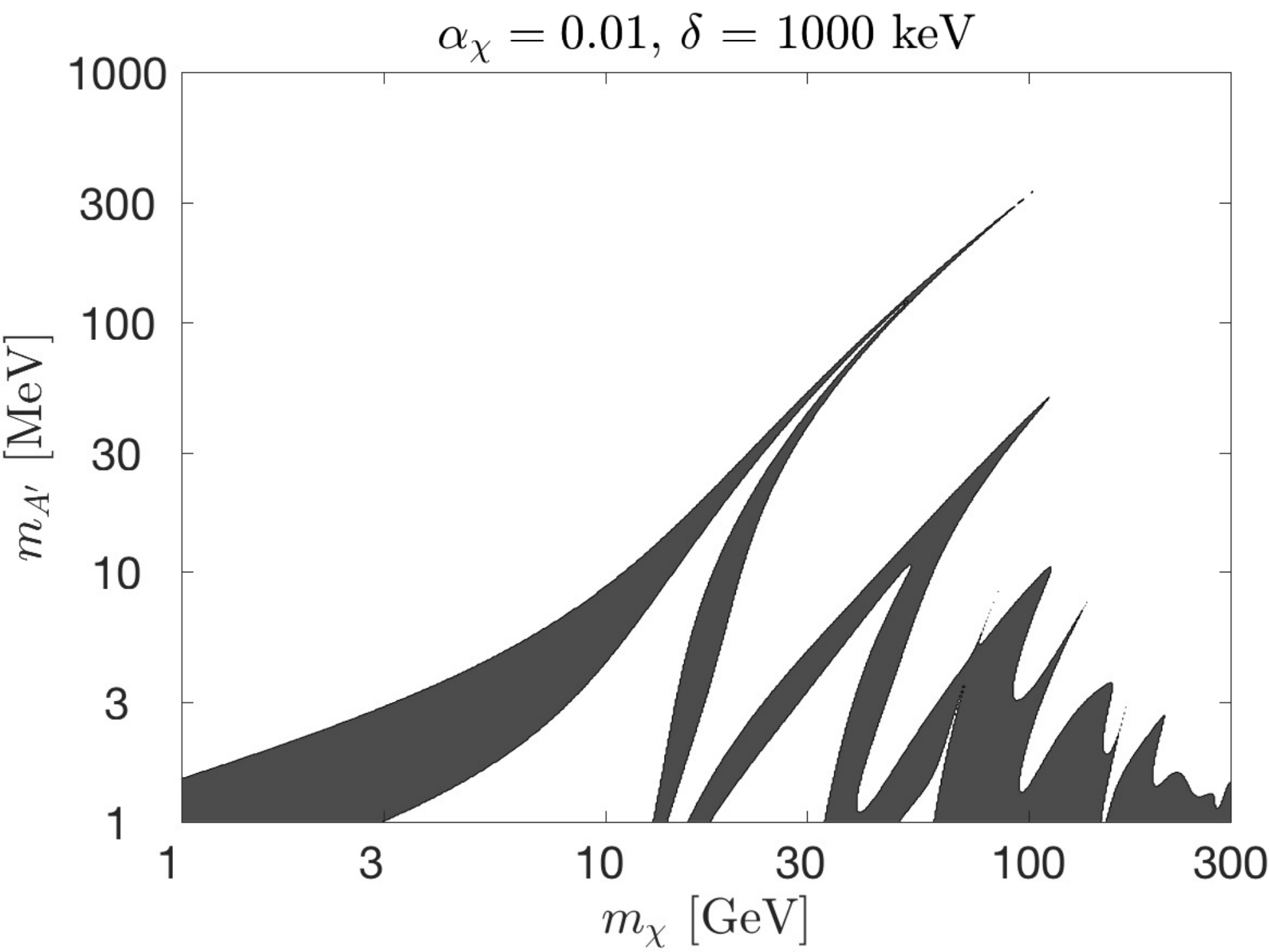}
	\caption{Greyed out is a region where $0.5<\sigma_{\chi \chi \rightarrow \chi \chi,V}<5$~cm$^2$/g is satisfied for $v = 10^{-4}c$ which is relevant to dwarf galaxies. The dark fine-structure constant is set to $\alpha_\chi = 0.01$ and the mass splitting is set to $\delta=150 \, (1000)$~keV in the left (right) plot.} \label{grid_comparison}
\end{figure}

Lastly, we present results using a fixed dark fine-structure constant $\alpha_\chi = 0.01$ and calculate the viscosity cross section for two cases, $\delta = 150$~keV and $1$~MeV. In order to have a reasonable computational time, we now fix the velocity at $10^{-4}c$. The results are shown in fig.~\ref{grid_comparison}. The same trend is visible where the cross-section decreases as the mass splitting is increased. We can directly compare the figure in which $\delta = 1$~MeV with ref.~\cite{Zhang:2016dck}. The main differences come from the fact that they calculate the transfer cross-section. Since the weight differs by an $\mathcal{O}(1)$ number for each partial wave, we expect overlapping regions and this is precisely what is seen. We also calculated the scattering cross section as a function of velocity for $\delta = 1$~MeV with $\alpha_\chi = 0.01$, and the results are essentially the same as for smaller $\delta$, see fig.~\ref{sigma_examples}. The only main difference is that the upscattering cross section is kinematically unlocked at a larger velocity, and that downscattering may be significant since $\alpha_\chi$ is much larger at the smaller DM masses.

\section{Abundance of the excited state} \label{abundance}
It is crucial to understand the DM composition of a typical halo. Next we will discuss why there may be no excited states $\chi^*$ left today, even if the $\chi^*$ is stable. The key point is that the same large self-scattering cross sections may drive the abundance of $\chi^*$ down to a negligible amount~\cite{Finkbeiner:2009mi,Batell:2009vb}.

If the DM are thermal relics, the temperature at which the DM production stops implies that there will be equal numbers of both species. If the excited state is cosmologically unstable, halos will naturally be composed of only the lower state $\chi$. This may very well be the case if $\delta > m_{A'}$, since then the decay $\chi^* \rightarrow \chi + A'$ is open and it is not suppressed by the mixing. If $\chi^*$ is cosmologically stable, its relic density may still be depleted by the self-scattering processes $\chi^*\chi^*\rightarrow \chi\chi$ in the early Universe. In order for there to be a large population of $\chi^*$, these processes must become inefficient at temperatures larger than the mass splitting $\delta$ such that the $\chi^*$ population is not Boltzmann suppressed. Taking $T_\chi$ to be the temperature of the DM particles and $T_\gamma$ to be the temperature of the SM particles, we can now estimate the temperature $T_\chi^{\rm dd}$ at which the double de-excitation rate drops below the Hubble rate\footnote{A more precise treatment would require solutions to the coupled Boltzmann equations in order to compute the density of the $\chi,\,\chi^*$ states, but for our purposes the following discussion is sufficient.}. That is, the inequality
\begin{equation} \label{T_dd}
\Gamma_{\chi^* \chi^*}(T_\chi) \equiv \left[\langle \sigma_{\chi^* \chi^* \rightarrow \chi \chi} v \rangle n_{\rm \chi^*}\right]_{T_\chi}<H(T_\gamma)
\end{equation}
holds at temperatures $T_\chi < T_\chi^{\rm dd}$. The ratio between the number densities of the excited and the lower mass state in thermal equilibrium is to a good accuracy given by $n_{\chi^*}(T_\gamma) = n_\chi(T_\gamma) e^{-\delta / T_\chi}$. After expressing $n_\chi$ in terms of the total DM density $n_{\rm tot}=n_\chi+n_{\chi^*}$, this can be rearranged to
\begin{equation}
n_{\chi^*}(T_\gamma) = \frac{1}{e^{\delta/T_\chi} + 1} n_{\rm tot}(T_\gamma)\,.
\end{equation}
The evolution of the total number density can be described in terms of the photon temperature at some higher temperature $T_\gamma$ and the current photon temperature $T_0$ as
\begin{equation}
n_{\rm tot}(T_\gamma) = \Omega_\chi \frac{\rho_{\rm crit}}{m_\chi} \frac{T^3_{\rm \gamma}}{T^3_0}\,,
\end{equation}
where $\Omega_\chi = 0.26$ is the DM relative abundance, $\rho_{\rm crit} = 3H_0^2 M^2_p/(8\pi)$ is the critical density today~\footnote{$\rho_{\rm crit} = 3.7\,[4.3] \cdot 10^{-47}$~GeV$^4$ for $H_0=67.80\,[73.00]\,{\rm km/s/Mpc}$ as determined by the Planck satellite~\cite{Ade:2015xua} [the Hubble Space Telescope~\cite{Riess:2016jrr}]. We use the Planck satellite result~\cite{Ade:2015xua}.} and $T_0 = 0.235$~meV is the CMB temperature today as measured by the Planck satellite~\cite{Ade:2015xua}. At some temperature $T_*$, the DM particles will kinetically decouple from the radiation bath which keeps it at the same temperature as the photons. Since this occurs when they are non-relativistic, their temperature will evolve as the inverse of the scale factor squared after this time. The relation $T_\chi = T^2_\gamma/T_*$ can then be used to relate the temperature of the DM particles and the photons. Assuming that freeze-out of de-excitations occurs in the radiation dominated period, we have
\begin{equation}
H(T) = \frac{1.66 \sqrt{g_*}}{m_{\rm pl}} T_\gamma^2\,,
\end{equation}
where $g_* \sim 10$ is the number of relativistic degrees of freedom at the time. The freeze-out temperature is then calculated by solving eq.~\eqref{T_dd} and subsequently the fractional abundance is found from
\begin{equation}
\frac{n_{\chi^*}}{n_\chi + n_{\chi^*}} = \frac{1}{e^{\delta/T_{\chi}^{\rm dd}}+1}\,.
\end{equation}
\begin{figure}
	\centering
	\includegraphics[width=0.48\textwidth]{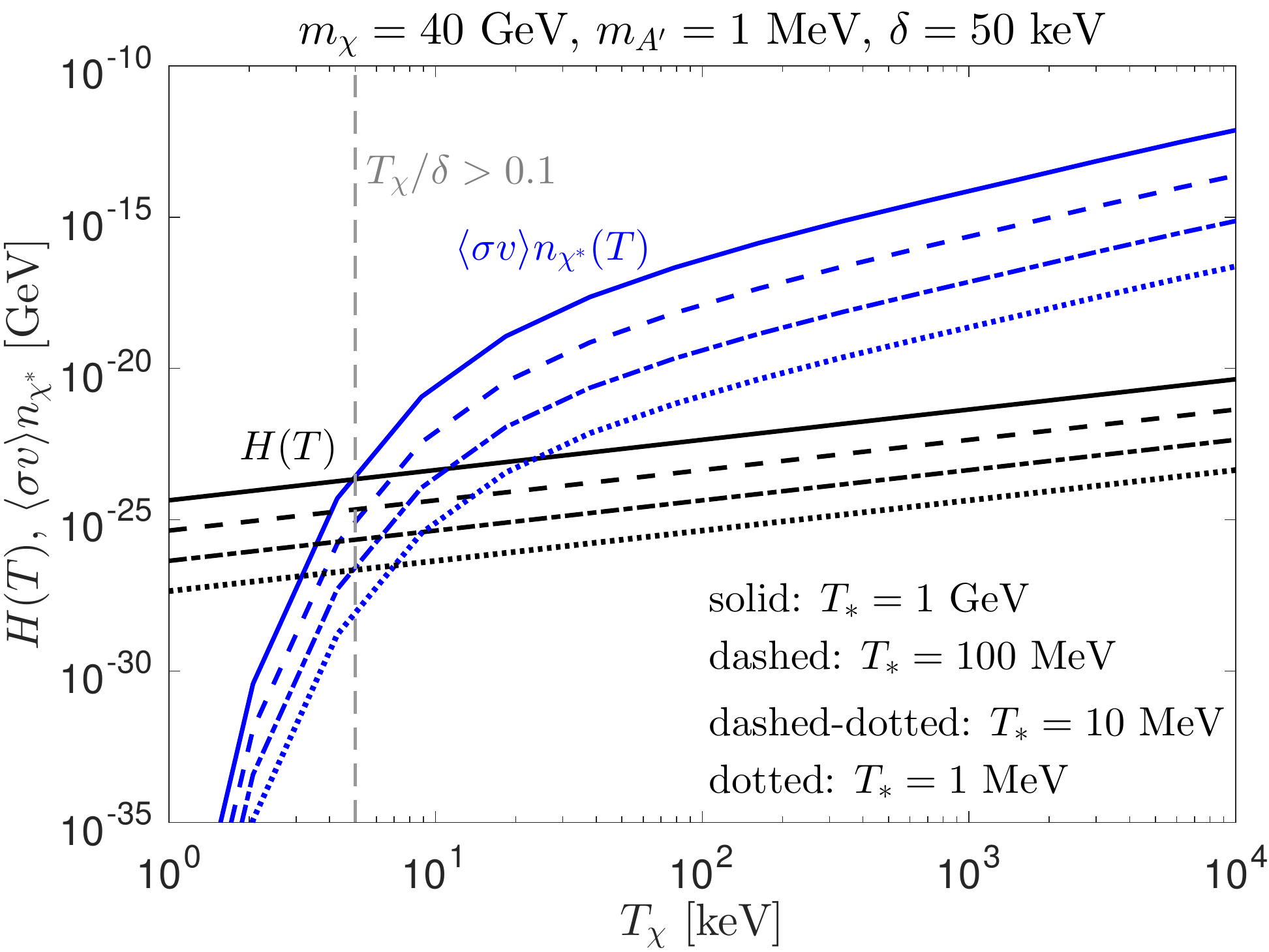}
	\includegraphics[width=0.48\textwidth]{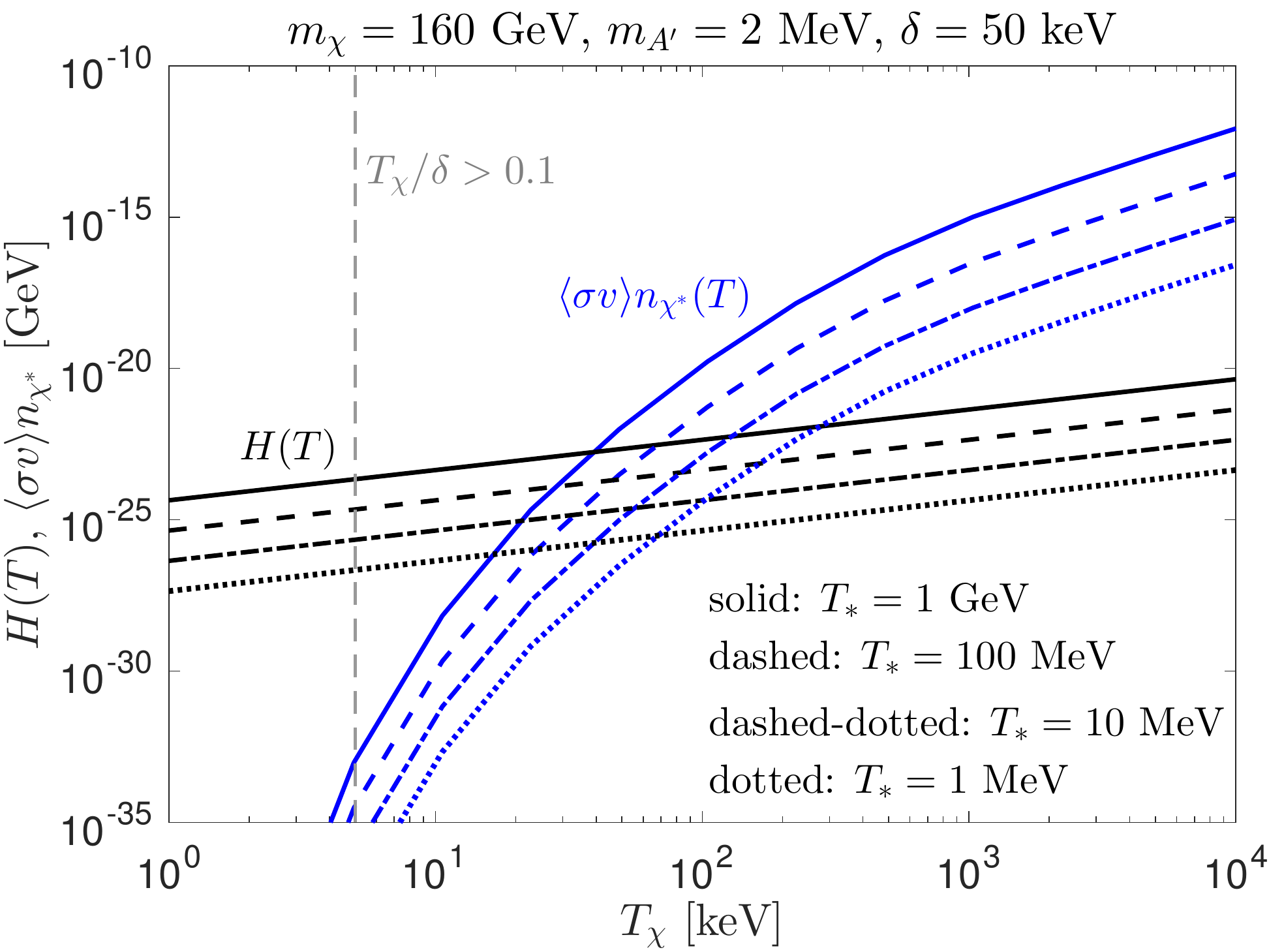}
	\caption{Comparison between the Hubble rate and the rate of the double de-excitation reaction $\chi^*\chi^* \rightarrow \chi\chi $ as a function of DM temperature for different values of the kinetic decoupling temperature, $T_*=1,\,10,\,100,\,1000$ MeV.} \label{relic_examples}
\end{figure}

Two examples of comparisons between $H(T)$ and $\Gamma_{\chi^* \chi^*}(T)$ are shown in fig.~\ref{relic_examples}. The scattering cross section is taken from benchmarks A and D for $\delta = 50$~keV (see table~\ref{benchmark_points}), which are also plotted in fig.~\ref{sigma_examples}. Below $T_\chi=\delta$, the de-excitation rate drops exponentially along with the abundance of the excited state. Generally, the longer kinetic equilibrium is kept, the higher $T_{\chi}^{\rm dd}$ will be. This is expected, since the DM temperature drops faster if it is not in equilibrium with the radiation bath. For the lower DM mass (left figure), $T_\chi^{\rm dd}$ is smaller than $\delta$ in all cases implying that the vast majority of DM is in the lower mass state. On the other hand, for the larger DM masses the interaction rate falls below the Hubble rate at a larger temperature that may allow for a significant ($\mathcal{O}(0.1)$) fraction of excited state particles. Also for the cases where $\delta = 10$~keV, the largest DM / smallest mediator masses give rise to the largest fraction of excited state particles which is also $\mathcal{O}(0.1)$ in the most extreme cases. This means that, indeed, direct detection signals are expected to be suppressed with respect to the elastic case.

\section{Discussion and conclusions} \label{conc}

We have studied self-interactions of a two-component inelastic DM model in the context of solving small scale structure problems. Of special interest is the part of parameter space where the DM has a weak scale mass and the mass splitting is in the range needed to evade or significantly relax direct detection experiments ($\delta \gtrsim \mathcal{O}(50)$ keV), and therefore be compatible with BBN constraints that apply to the decay of the mediator. We also discuss to what extent the DM composition of a halo contains the excited state.

We find several interesting features for the self-scattering cross section. There are large regions in which the small scale structure problems can be solved by elastic scattering cross sections. Self-scattering amplitudes between two $\chi$ or $\chi^*$ are very similar in magnitude in the range of velocities interesting in the dwarf and LSB galaxy halos allowing for both components to contribute to the formation of cores. The de-excitation amplitude for the excited state at velocities where up-scattering is kinematically forbidden can be reduced by many orders of magnitude with respect to elastic scattering. It's worth pointing out that comparing to the case of self-scattering of elastic DM, generally the mediator is required to have a smaller mass to achieve large enough self-scattering cross sections.

Even if the excited state is stable on cosmological time scales, it is expected that the abundance of the excited state is subdominant due to the large self-interactions required. The probability for a non-negligible fraction of excited state particles to survive requires heavier DM and/or mediator mass to suppress the de-excitation cross section, and/or that the dark sector is kept in kinetic equilibrium for as long as possible. This should be properly analyzed for a specific model.

The large suppression of the de-excitation cross section also protects the halo from evaporation. Since the solution to small scale structures requires the average scattering rate per particle to be of $\mathcal{O}(1)$ over the lifetime of the halo in question, the average scattering rate for de-excitations of two $\chi^*$ will be orders of magnitude less as long as the DM is heavy enough. Given a down-scattering cross section orders of magnitude smaller, evaporation would be negligible for a halo with a sizeable $\chi^*$ population even if every particle that down-scatters escapes from the halo.

All in all, inelastic scattering seems like a very simple and natural scenario for large self-interactions in order to solve small-scale structure problems, while at the same time being able to evade direct detection limits. If 1) small-scale structure problems in the cold DM paradigm persist, 2) next-generation DD experiments set stronger upper limits, and 3) no signals are observed from dwarf galaxies or the CMB, a search for models of inelastic DM with scalar mediators and p-wave annihilation cross-sections should be pursued.

\vspace{0.5cm}
{\bf Note added:} During the final stages of this work, a paper on inelastic self-interacting dark matter appeared on arXiv~\cite{Zhang:2016dck}. Their solution of the coupled Schr\"odinger equation uses an adiabatic approximation which requires very large splittings and small $\alpha_\chi$. They also do not consider interactions amongst the excited state particles. Comparing our solution to theirs, we find good agreement (see fig.~\ref{grid_comparison} right).

\section*{Acknowledgments}

We want to give our warmest thanks Bryan Zaldivar for his participation in the initial stages of this project and for many illuminating discussions. This work was supported by the G\"oran Gustafsson foundation. JHG acknowledges the support from the University of Adelaide and the Australian Research Council through the ARC Centre of Excellence for Particle Physics at the Terascale (CoEPP) (CE110001104).

\bibliographystyle{my-h-physrev}
\bibliography{refs}

\begin{thebibliography}{10}

\bibitem{Navarro:1995iw}
J.~F. Navarro, C.~S. Frenk, and S.~D.~M. White,
\newblock {\em {The Structure of cold dark matter halos}},
\newblock Astrophys. J. {\bf 462}, 563 (1996), astro-ph/9508025.

\bibitem{Moore:1997sg}
B.~Moore, F.~Governato, T.~R. Quinn, J.~Stadel, and G.~Lake,
\newblock {\em {Resolving the structure of cold dark matter halos}},
\newblock Astrophys. J. {\bf 499}, L5 (1998), astro-ph/9709051.

\bibitem{Klypin:2000hk}
A.~Klypin, A.~V. Kravtsov, J.~Bullock, and J.~Primack,
\newblock {\em {Resolving the structure of cold dark matter halos}},
\newblock Astrophys. J. {\bf 554}, 903 (2001), astro-ph/0006343.

\bibitem{Diemand:2005wv}
J.~Diemand, M.~Zemp, B.~Moore, J.~Stadel, and M.~Carollo,
\newblock {\em {Cusps in cold dark matter haloes}},
\newblock Mon. Not. Roy. Astron. Soc. {\bf 364}, 665 (2005), astro-ph/0504215.

\bibitem{Springel:2008cc}
V.~Springel {\em et~al.},
\newblock {\em {The Aquarius Project: the subhalos of galactic halos}},
\newblock Mon. Not. Roy. Astron. Soc. {\bf 391}, 1685 (2008), 0809.0898.

\bibitem{Moore:1994yx}
B.~Moore,
\newblock {\em {Evidence against dissipationless dark matter from observations
  of galaxy haloes}},
\newblock Nature {\bf 370}, 629 (1994).

\bibitem{Flores:1994gz}
R.~A. Flores and J.~R. Primack,
\newblock {\em {Observational and theoretical constraints on singular dark
  matter halos}},
\newblock Astrophys. J. {\bf 427}, L1 (1994), astro-ph/9402004.

\bibitem{deBlok:2001hbg}
W.~J.~G. de~Blok, S.~S. McGaugh, A.~Bosma, and V.~C. Rubin,
\newblock {\em {Mass density profiles of LSB galaxies}},
\newblock Astrophys. J. {\bf 552}, L23 (2001), astro-ph/0103102.

\bibitem{deBlok:2002vgq}
W.~J.~G. de~Blok and A.~Bosma,
\newblock {\em {High-resolution rotation curves of low surface brightness
  galaxies}},
\newblock Astron. Astrophys. {\bf 385}, 816 (2002), astro-ph/0201276.

\bibitem{Swaters:2002rx}
R.~A. Swaters, B.~F. Madore, F.~C. van~den Bosch, and M.~Balcells,
\newblock {\em {The Central mass distribution in dwarf and low-surface
  brightness galaxies}},
\newblock Astrophys. J. {\bf 583}, 732 (2003), astro-ph/0210152.

\bibitem{Simon:2004sr}
J.~D. Simon, A.~D. Bolatto, A.~Leroy, L.~Blitz, and E.~L. Gates,
\newblock {\em {High-resolution measurements of the halos of four dark
  matter-dominated galaxies: Deviations from a universal density profile}},
\newblock Astrophys. J. {\bf 621}, 757 (2005), astro-ph/0412035.

\bibitem{Spekkens:2005ik}
K.~Spekkens and R.~Giovanelli,
\newblock {\em {The Cusp/core problem in Galactic halos: Long-slit spectra for
  a large dwarf galaxy sample}},
\newblock Astron. J. {\bf 129}, 2119 (2005), astro-ph/0502166.

\bibitem{KuziodeNaray:2007qi}
R.~Kuzio~de Naray, S.~S. McGaugh, and W.~J.~G. de~Blok,
\newblock {\em {Mass Models for Low Surface Brightness Galaxies with High
  Resolution Optical Velocity Fields}},
\newblock Astrophys. J. {\bf 676}, 920 (2008), 0712.0860.

\bibitem{Spano:2007nt}
M.~Spano {\em et~al.},
\newblock {\em {GHASP: An H-alpha kinematic survey of spiral and irregular
  galaxies. 5. Dark matter distribution in 36 nearby spiral galaxies}},
\newblock Mon. Not. Roy. Astron. Soc. {\bf 383}, 297 (2008), 0710.1345.

\bibitem{deBlok:2008wp}
W.~J.~G. de~Blok {\em et~al.},
\newblock {\em {High-Resolution Rotation Curves and Galaxy Mass Models from
  THINGS}},
\newblock Astron. J. {\bf 136}, 2648 (2008), 0810.2100.

\bibitem{Newman:2009qm}
A.~B. Newman {\em et~al.},
\newblock {\em {The Distribution of Dark Matter Over 3 Decades in Radius in the
  Lensing Cluster Abell 611}},
\newblock Astrophys. J. {\bf 706}, 1078 (2009), 0909.3527.

\bibitem{Donato:2009ab}
F.~Donato {\em et~al.},
\newblock {\em {A constant dark matter halo surface density in galaxies}},
\newblock Mon. Not. Roy. Astron. Soc. {\bf 397}, 1169 (2009), 0904.4054.

\bibitem{Oh:2010ea}
S.-H. Oh, W.~J.~G. de~Blok, E.~Brinks, F.~Walter, and R.~C. Kennicutt, Jr,
\newblock {\em {Dark and luminous matter in THINGS dwarf galaxies}},
\newblock Astron. J. {\bf 141}, 193 (2011), 1011.0899.

\bibitem{Walker:2011zu}
M.~G. Walker and J.~Penarrubia,
\newblock {\em {A Method for Measuring (Slopes of) the Mass Profiles of Dwarf
  Spheroidal Galaxies}},
\newblock Astrophys. J. {\bf 742}, 20 (2011), 1108.2404.

\bibitem{Newman:2012nw}
A.~B. Newman, T.~Treu, R.~S. Ellis, and D.~J. Sand,
\newblock {\em {The Density Profiles of Massive, Relaxed Galaxy Clusters: II.
  Separating Luminous and Dark Matter in Cluster Cores}},
\newblock Astrophys. J. {\bf 765}, 25 (2013), 1209.1392.

\bibitem{Adams:2014bda}
J.~J. Adams {\em et~al.},
\newblock {\em {Dwarf Galaxy Dark Matter Density Profiles Inferred from Stellar
  and Gas Kinematics}},
\newblock Astrophys. J. {\bf 789}, 63 (2014), 1405.4854.

\bibitem{Spergel:1999mh}
D.~N. Spergel and P.~J. Steinhardt,
\newblock {\em {Observational evidence for selfinteracting cold dark matter}},
\newblock Phys. Rev. Lett. {\bf 84}, 3760 (2000), astro-ph/9909386.

\bibitem{Firmani:2000ce}
C.~Firmani, E.~D'Onghia, V.~Avila-Reese, G.~Chincarini, and X.~Hernandez,
\newblock {\em {Evidence of self-interacting cold dark matter from galactic to
  galaxy cluster scales}},
\newblock Mon. Not. Roy. Astron. Soc. {\bf 315}, L29 (2000), astro-ph/0002376.

\bibitem{Hu:2000ke}
W.~Hu, R.~Barkana, and A.~Gruzinov,
\newblock {\em {Cold and fuzzy dark matter}},
\newblock Phys. Rev. Lett. {\bf 85}, 1158 (2000), astro-ph/0003365.

\bibitem{Goodman:2000tg}
J.~Goodman,
\newblock {\em {Repulsive dark matter}},
\newblock New Astron. {\bf 5}, 103 (2000), astro-ph/0003018.

\bibitem{Marsh:2013ywa}
D.~J.~E. Marsh and J.~Silk,
\newblock {\em {A Model For Halo Formation With Axion Mixed Dark Matter}},
\newblock Mon. Not. Roy. Astron. Soc. {\bf 437}, 2652 (2014), 1307.1705.

\bibitem{Schive:2014dra}
H.-Y. Schive, T.~Chiueh, and T.~Broadhurst,
\newblock {\em {Cosmic Structure as the Quantum Interference of a Coherent Dark
  Wave}},
\newblock Nature Phys. {\bf 10}, 496 (2014), 1406.6586.

\bibitem{Schive:2014hza}
H.-Y. Schive {\em et~al.},
\newblock {\em {Understanding the Core-Halo Relation of Quantum Wave Dark
  Matter from 3D Simulations}},
\newblock Phys. Rev. Lett. {\bf 113}, 261302 (2014), 1407.7762.

\bibitem{Marsh:2015wka}
D.~J.~E. Marsh and A.-R. Pop,
\newblock {\em {Axion dark matter, solitons and the cusp--core problem}},
\newblock Mon. Not. Roy. Astron. Soc. {\bf 451}, 2479 (2015), 1502.03456.

\bibitem{Dave:2000ar}
R.~Dave, D.~N. Spergel, P.~J. Steinhardt, and B.~D. Wandelt,
\newblock {\em {Halo properties in cosmological simulations of selfinteracting
  cold dark matter}},
\newblock Astrophys. J. {\bf 547}, 574 (2001), astro-ph/0006218.

\bibitem{Zavala:2012us}
J.~Zavala, M.~Vogelsberger, and M.~G. Walker,
\newblock {\em {Constraining Self-Interacting Dark Matter with the Milky Way's
  dwarf spheroidals}},
\newblock Monthly Notices of the Royal Astronomical Society: Letters {\bf 431},
  L20 (2013), 1211.6426.

\bibitem{Rocha:2012jg}
M.~Rocha {\em et~al.},
\newblock {\em {Cosmological Simulations with Self-Interacting Dark Matter I:
  Constant Density Cores and Substructure}},
\newblock Mon. Not. Roy. Astron. Soc. {\bf 430}, 81 (2013), 1208.3025.

\bibitem{Elbert:2014bma}
O.~D. Elbert {\em et~al.},
\newblock {\em {Core formation in dwarf haloes with self-interacting dark
  matter: no fine-tuning necessary}},
\newblock Mon. Not. Roy. Astron. Soc. {\bf 453}, 29 (2015), 1412.1477.

\bibitem{Vogelsberger:2015gpr}
M.~Vogelsberger {\em et~al.},
\newblock {\em {ETHOS - An Effective Theory of Structure Formation: Dark matter
  physics as a possible explanation of the small-scale CDM problems}},
\newblock Mon. Not. Roy. Astron. Soc. {\bf 460}, 1399 (2016), 1512.05349.

\bibitem{Vogelsberger:2012ku}
M.~Vogelsberger, J.~Zavala, and A.~Loeb,
\newblock {\em {Subhaloes in Self-Interacting Galactic Dark Matter Haloes}},
\newblock Mon. Not. Roy. Astron. Soc. {\bf 423}, 3740 (2012), 1201.5892.

\bibitem{Fry:2015rta}
A.~B. Fry {\em et~al.},
\newblock {\em {All about baryons: revisiting SIDM predictions at small halo
  masses}},
\newblock Mon. Not. Roy. Astron. Soc. {\bf 452}, 1468 (2015), 1501.00497.

\bibitem{Harvey:2015hha}
D.~Harvey, R.~Massey, T.~Kitching, A.~Taylor, and E.~Tittley,
\newblock {\em {The non-gravitational interactions of dark matter in colliding
  galaxy clusters}},
\newblock Science {\bf 347}, 1462 (2015), 1503.07675.

\bibitem{Randall:2007ph}
S.~W. Randall, M.~Markevitch, D.~Clowe, A.~H. Gonzalez, and M.~Bradac,
\newblock {\em {Constraints on the Self-Interaction Cross-Section of Dark
  Matter from Numerical Simulations of the Merging Galaxy Cluster 1E 0657-56}},
\newblock Astrophys. J. {\bf 679}, 1173 (2008), 0704.0261.

\bibitem{Peter:2012jh}
A.~H.~G. Peter, M.~Rocha, J.~S. Bullock, and M.~Kaplinghat,
\newblock {\em {Cosmological Simulations with Self-Interacting Dark Matter II:
  Halo Shapes vs. Observations}},
\newblock Mon. Not. Roy. Astron. Soc. {\bf 430}, 105 (2013), 1208.3026.

\bibitem{Loeb:2010gj}
A.~Loeb and N.~Weiner,
\newblock {\em {Cores in Dwarf Galaxies from Dark Matter with a Yukawa
  Potential}},
\newblock Phys. Rev. Lett. {\bf 106}, 171302 (2011), 1011.6374.

\bibitem{Tulin:2013teo}
S.~Tulin, H.-B. Yu, and K.~M. Zurek,
\newblock {\em {Beyond Collisionless Dark Matter: Particle Physics Dynamics for
  Dark Matter Halo Structure}},
\newblock Phys. Rev. {\bf D87}, 115007 (2013), 1302.3898.

\bibitem{Kaplinghat:2015aga}
M.~Kaplinghat, S.~Tulin, and H.-B. Yu,
\newblock {\em {Dark Matter Halos as Particle Colliders: Unified Solution to
  Small-Scale Structure Puzzles from Dwarfs to Clusters}},
\newblock Phys. Rev. Lett. {\bf 116}, 041302 (2016), 1508.03339.

\bibitem{Dent:2012mx}
J.~B. Dent, F.~Ferrer, and L.~M. Krauss,
\newblock {\em {Constraints on Light Hidden Sector Gauge Bosons from Supernova
  Cooling}},
\newblock (2012), 1201.2683.

\bibitem{Kazanas:2014mca}
D.~Kazanas, R.~N. Mohapatra, S.~Nussinov, V.~L. Teplitz, and Y.~Zhang,
\newblock {\em {Supernova Bounds on the Dark Photon Using its Electromagnetic
  Decay}},
\newblock Nucl. Phys. {\bf B890}, 17 (2014), 1410.0221.

\bibitem{Rrapaj:2015wgs}
E.~Rrapaj and S.~Reddy,
\newblock {\em {Nucleon-nucleon bremsstrahlung of dark gauge bosons and revised
  supernova constraints}},
\newblock Phys. Rev. {\bf C94}, 045805 (2016), 1511.09136.

\bibitem{Chang:2016ntp}
J.~H. Chang, R.~Essig, and S.~D. McDermott,
\newblock {\em {Revisiting Supernova 1987A Constraints on Dark Photons}},
\newblock (2016), 1611.03864.

\bibitem{Hardy:2016kme}
E.~Hardy and R.~Lasenby,
\newblock {\em {Stellar cooling bounds on new light particles: including plasma
  effects}},
\newblock (2016), 1611.05852.

\bibitem{Kaplinghat:2013yxa}
M.~Kaplinghat, S.~Tulin, and H.-B. Yu,
\newblock {\em {Direct Detection Portals for Self-interacting Dark Matter}},
\newblock Phys. Rev. {\bf D89}, 035009 (2014), 1310.7945.

\bibitem{Batell:2009vb}
B.~Batell, M.~Pospelov, and A.~Ritz,
\newblock {\em {Direct Detection of Multi-component Secluded WIMPs}},
\newblock Phys. Rev. {\bf D79}, 115019 (2009), 0903.3396.

\bibitem{DelNobile:2015uua}
E.~Del~Nobile, M.~Kaplinghat, and H.-B. Yu,
\newblock {\em {Direct Detection Signatures of Self-Interacting Dark Matter
  with a Light Mediator}},
\newblock JCAP {\bf 1510}, 055 (2015), 1507.04007.

\bibitem{Bringmann:2016din}
T.~Bringmann, F.~Kahlhoefer, K.~Schmidt-Hoberg, and P.~Walia,
\newblock {\em {Strong constraints on self-interacting dark matter with light
  mediators}},
\newblock (2016), 1612.00845.

\bibitem{TuckerSmith:2001hy}
D.~Tucker-Smith and N.~Weiner,
\newblock {\em {Inelastic dark matter}},
\newblock Phys. Rev. {\bf D64}, 043502 (2001), hep-ph/0101138.

\bibitem{TuckerSmith:2004jv}
D.~Tucker-Smith and N.~Weiner,
\newblock {\em {The Status of inelastic dark matter}},
\newblock Phys. Rev. {\bf D72}, 063509 (2005), hep-ph/0402065.

\bibitem{Chang:2008gd}
S.~Chang, G.~D. Kribs, D.~Tucker-Smith, and N.~Weiner,
\newblock {\em {Inelastic Dark Matter in Light of DAMA/LIBRA}},
\newblock Phys. Rev. {\bf D79}, 043513 (2009), 0807.2250.

\bibitem{Graham:2010ca}
P.~W. Graham, R.~Harnik, S.~Rajendran, and P.~Saraswat,
\newblock {\em {Exothermic Dark Matter}},
\newblock Phys. Rev. {\bf D82}, 063512 (2010), 1004.0937.

\bibitem{An:2011uq}
H.~An, P.~S.~B. Dev, Y.~Cai, and R.~N. Mohapatra,
\newblock {\em {Sneutrino Dark Matter in Gauged Inverse Seesaw Models for
  Neutrinos}},
\newblock Phys. Rev. Lett. {\bf 108}, 081806 (2012), 1110.1366.

\bibitem{Schwetz:2011xm}
T.~Schwetz and J.~Zupan,
\newblock {\em {Dark Matter attempts for CoGeNT and DAMA}},
\newblock JCAP {\bf 1108}, 008 (2011), 1106.6241.

\bibitem{McCullough:2013jma}
M.~McCullough and L.~Randall,
\newblock {\em {Exothermic Double-Disk Dark Matter}},
\newblock JCAP {\bf 1310}, 058 (2013), 1307.4095.

\bibitem{Fox:2013pia}
P.~J. Fox, G.~Jung, P.~Sorensen, and N.~Weiner,
\newblock {\em {Dark matter in light of the LUX results}},
\newblock Phys. Rev. {\bf D89}, 103526 (2014), 1401.0216.

\bibitem{Bozorgnia:2013hsa}
N.~Bozorgnia, J.~Herrero-Garcia, T.~Schwetz, and J.~Zupan,
\newblock {\em {Halo-independent methods for inelastic dark matter
  scattering}},
\newblock JCAP {\bf 1307}, 049 (2013), 1305.3575.

\bibitem{Frandsen:2014ima}
M.~T. Frandsen and I.~M. Shoemaker,
\newblock {\em {Up-shot of inelastic down-scattering at CDMS-Si}},
\newblock Phys. Rev. {\bf D89}, 051701 (2014), 1401.0624.

\bibitem{Chen:2014tka}
N.~Chen {\em et~al.},
\newblock {\em {Exothermic isospin-violating dark matter after SuperCDMS and
  CDEX}},
\newblock Phys. Lett. {\bf B743}, 205 (2015), 1404.6043.

\bibitem{Blennow:2015hzp}
M.~Blennow, S.~Clementz, and J.~Herrero-Garcia,
\newblock {\em {Pinning down inelastic dark matter in the Sun and in direct
  detection}},
\newblock JCAP {\bf 1604}, 004 (2016), 1512.03317.

\bibitem{Tan:2016zwf}
PandaX-II, A.~Tan {\em et~al.},
\newblock {\em {Dark Matter Results from First 98.7 Days of Data from the
  PandaX-II Experiment}},
\newblock Phys. Rev. Lett. {\bf 117}, 121303 (2016), 1607.07400.

\bibitem{Akerib:2015rjg}
LUX, D.~S. Akerib {\em et~al.},
\newblock {\em {Improved Limits on Scattering of Weakly Interacting Massive
  Particles from Reanalysis of 2013 LUX Data}},
\newblock Phys. Rev. Lett. {\bf 116}, 161301 (2016), 1512.03506.

\bibitem{Agnese:2015nto}
SuperCDMS, R.~Agnese {\em et~al.},
\newblock {\em {New Results from the Search for Low-Mass Weakly Interacting
  Massive Particles with the CDMS Low Ionization Threshold Experiment}},
\newblock Phys. Rev. Lett. {\bf 116}, 071301 (2016), 1509.02448.

\bibitem{Schutz:2014nka}
K.~Schutz and T.~R. Slatyer,
\newblock {\em {Self-Scattering for Dark Matter with an Excited State}},
\newblock JCAP {\bf 1501}, 021 (2015), 1409.2867.

\bibitem{Zhang:2016dck}
Y.~Zhang,
\newblock {\em {Self-interacting Dark Matter Without Direct Detection
  Constraints}},
\newblock Phys. Dark Univ. {\bf 15}, 82 (2017), 1611.03492.

\bibitem{Boddy:2016bbu}
K.~K. Boddy, M.~Kaplinghat, A.~Kwa, and A.~H.~G. Peter,
\newblock {\em {Hidden Sector Hydrogen as Dark Matter: Small-scale Structure
  Formation Predictions and the Importance of Hyperfine Interactions}},
\newblock (2016), 1609.03592.

\bibitem{SanchezSalcedo:2003pb}
F.~J. Sanchez-Salcedo,
\newblock {\em {Unstable cold dark matter and the cuspy halo problem in dwarf
  galaxies}},
\newblock Astrophys. J. {\bf 591}, L107 (2003), astro-ph/0305496.

\bibitem{Abdelqader:2008wa}
M.~Abdelqader and F.~Melia,
\newblock {\em {Decaying Dark Matter and the Deficit of Dwarf Haloes}},
\newblock Mon. Not. Roy. Astron. Soc. {\bf 388}, 1869 (2008), 0806.0602.

\bibitem{Peter:2010au}
A.~H.~G. Peter,
\newblock {\em {Mapping the allowed parameter space for decaying dark matter
  models}},
\newblock Phys. Rev. {\bf D81}, 083511 (2010), 1001.3870.

\bibitem{Peter:2010jy}
A.~H.~G. Peter, C.~E. Moody, and M.~Kamionkowski,
\newblock {\em {Dark-Matter Decays and Self-Gravitating Halos}},
\newblock Phys. Rev. {\bf D81}, 103501 (2010), 1003.0419.

\bibitem{Peter:2010sz}
A.~H.~G. Peter and A.~J. Benson,
\newblock {\em {Dark-matter decays and Milky Way satellite galaxies}},
\newblock Phys. Rev. {\bf D82}, 123521 (2010), 1009.1912.

\bibitem{Bell:2010qt}
N.~F. Bell, A.~J. Galea, and R.~R. Volkas,
\newblock {\em {A Model For Late Dark Matter Decay}},
\newblock Phys. Rev. {\bf D83}, 063504 (2011), 1012.0067.

\bibitem{Cui:2009xq}
Y.~Cui, D.~E. Morrissey, D.~Poland, and L.~Randall,
\newblock {\em {Candidates for Inelastic Dark Matter}},
\newblock JHEP {\bf 05}, 076 (2009), 0901.0557.

\bibitem{ArkaniHamed:2008qn}
N.~Arkani-Hamed, D.~P. Finkbeiner, T.~R. Slatyer, and N.~Weiner,
\newblock {\em {A Theory of Dark Matter}},
\newblock Phys. Rev. {\bf D79}, 015014 (2009), 0810.0713.

\bibitem{Finkbeiner:2008gw}
D.~P. Finkbeiner, N.~Padmanabhan, and N.~Weiner,
\newblock {\em {CMB and 21-cm Signals for Dark Matter with a Long-Lived Excited
  State}},
\newblock Phys. Rev. {\bf D78}, 063530 (2008), 0805.3531.

\bibitem{Finkbeiner:2009mi}
D.~P. Finkbeiner, T.~R. Slatyer, N.~Weiner, and I.~Yavin,
\newblock {\em {PAMELA, DAMA, INTEGRAL and Signatures of Metastable Excited
  WIMPs}},
\newblock JCAP {\bf 0909}, 037 (2009), 0903.1037.

\bibitem{Slatyer:2009vg}
T.~R. Slatyer,
\newblock {\em {The Sommerfeld enhancement for dark matter with an excited
  state}},
\newblock JCAP {\bf 1002}, 028 (2010), 0910.5713.

\bibitem{Bernal:2015ova}
N.~Bernal, X.~Chu, C.~Garcia-Cely, T.~Hambye, and B.~Zaldivar,
\newblock {\em {Production Regimes for Self-Interacting Dark Matter}},
\newblock JCAP {\bf 1603}, 018 (2016), 1510.08063.

\bibitem{Chu:2016pew}
X.~Chu, C.~Garcia-Cely, and T.~Hambye,
\newblock {\em {Can the relic density of self-interacting dark matter be due to
  annihilations into Standard Model particles?}},
\newblock JHEP {\bf 11}, 048 (2016), 1609.00399.

\bibitem{Ade:2015xua}
Planck, P.~A.~R. Ade {\em et~al.},
\newblock {\em {Planck 2015 results. XIII. Cosmological parameters}},
\newblock Astron. Astrophys. {\bf 594}, A13 (2016), 1502.01589.

\bibitem{Riess:2016jrr}
A.~G. Riess {\em et~al.},
\newblock {\em {A 2.4\% Determination of the Local Value of the Hubble
  Constant}},
\newblock Astrophys. J. {\bf 826}, 56 (2016), 1604.01424.

\bibitem{Ershov:2011zz}
S.~N. Ershov, J.~S. Vaagen, and M.~V. Zhukov,
\newblock {\em {Modified variable phase method for the solution of coupled
  radial Schrodinger equations}},
\newblock Phys. Rev. {\bf C84}, 064308 (2011).

\end{thebibliography}

\makeatother
\appendix
\section{Computation of the self-interacting cross sections} \label{sec:details}

We focus on scattering of two $\chi$ (or $\chi^*$) since the case of $\chi \chi^*$ scattering is dealt with by the method presented in ref.~\cite{Tulin:2013teo}. The scattered wave function at very large $r$ can be written as
\begin{equation}
\Psi = \Psi_{\rm in}e^{i k_{\rm in} z} +
\frac{1}{r}\left( \begin{array}{cc}
f_X(\theta) e^{ikr}\\
f_Y(\theta) e^{ik'r}
\end{array} \right),
\end{equation}
where we have neglected an overall normalization constant. $\Psi_{\rm in}$ will be $(1\,\,0)^T$, with $k_{\rm in}=k$ [$(0\,\,1)^T$, with $k_{\rm in}=k'$] for $\chi \chi$ [$\chi^* \chi^*$] scattering. Expanding the above in partial waves yields
\begin{equation} \label{wave_func}
\Psi = \sum_l (2l+1)P_l({\rm cos} \, \theta) \left[\Psi_{\rm in} \frac{ e^{ik_{\rm in}r} - (-1)^l e^{-ik_{\rm in}r}}{2ik_{\rm in}r} +
\frac{1}{r}\left( \begin{array}{cc}
f_{X,l} e^{ikr}\\
f_{Y,l} e^{ik'r}
\end{array} \right)\right].
\end{equation}
The differential cross sections are then given by
\begin{equation} \label{diff_cross_sec}
\frac{d\sigma}{d\Omega} = \frac{k_{\rm out}}{k_{\rm in}} \left| \sum_l (2l+1) P_l({\rm cos} \, \theta) f_l \right| ^2\,,
\end{equation}
where for $\chi \chi$ scattering $f_l=f_{X,l}$ gives the elastic cross section, $f_l=f_{Y,l}$ the inelastic one and $k_{\rm in}$ ($k_{\rm out}$) is the momentum of the incoming (outgoing) state. Similarly, for $\chi^* \chi^*$ scattering, $f_{Y,l}$ provides the elastic cross section and $f_l=f_{X,l}$ the down-scattering one.

To find the constants $f_{X,l}$ and $f_{Y,l}$, eq.~\eqref{partial_schr_eqn} has to be solved and mapped onto the form
\begin{equation} \label{chi_num}
R_l(x) = \frac{1}{r} \left( \begin{array}{cc}
A_l e^{-ikr} + B_l e^{ikr}\\
C_l e^{-ik'r} + D_l e^{ik'r}
\end{array} \right).
\end{equation}
For incoming $\chi$ ($\chi^*$) particles, $C_l=0$ ($A_l=0$). In the first case, multiplying the wave-function by $(-1)^{l+1}/(2ik_{\rm in}A_l)$ and comparing to eq.~\eqref{wave_func} allows for the identifications
\begin{equation}
\frac{(-1)^{l+1}}{2ik_{\rm in}} \frac{B_l}{A_l} = \frac{1}{2ik_{\rm in}} + f_{X,l}, \quad \frac{(-1)^{l+1}}{2ik_{\rm in}} \frac{D_l}{A_l} = f^{Y,l}.
\end{equation}
For $\chi^* \chi^*$ scattering, the following substitutions are in order: $f_{X,l} \longleftrightarrow f_{Y,l}$, $A_l \longleftrightarrow C_l$, $B_l \longleftrightarrow D_l$.

It is convenient to introduce the following change of variables
\begin{equation}
\quad x = 2 \alpha_\chi \mu r, \quad a = \frac{v}{2 \alpha_\chi}, \quad b = \frac{2 \alpha_\chi \mu}{m_{A'}}, \quad c^2 = -\frac{\delta}{\mu \alpha_\chi^2} + a^2,
\end{equation}
with which eq.~\eqref{wave_func} can be written as
\begin{equation} \label{wave_funcb}
\Psi = \sum_l (2l+1)P_l({\rm cos} \, \theta) \left[\Psi_{\rm in} \frac{ e^{ip_{\rm in}x} - (-1)^l e^{-ip_{\rm in}x}}{2ip_{\rm in}x} +
\frac{1}{x} \left( \begin{array}{cc}
f_{X,l} e^{iax}\\
f_{Y,l} e^{icx}
\end{array} \right)\right]
\end{equation}
where $p_{\rm in} = a$ for $\chi \chi$ scattering and $p_{\rm in} = c$ for $\chi^* \chi^*$ scattering. Note that $f_{X,l}$ and $f_{Y,l}$ should be rescaled by $\hbar / (\alpha_\chi m_\chi c_l)$ (where $c_l$ is the speed of light) when going back to $R_l(r)$ to compute the physical cross sections. Finally, the redefinition 
\begin{equation} \label{R_chi_link}
R_l(x) = \frac{\chi_l(x)}{x}
\end{equation}
turns eq.~\eqref{partial_schr_eqn} into
\begin{equation} \label{schrodinger_eq}
\chi''_l = \left(
\begin{array}{cc}
\frac{l(l+1)}{x^2} - a^2 & -\frac{1}{x}e^{-x/b} \\
-\frac{1}{x}e^{-x/b} & \frac{l(l+1)}{x^2} - c^2
\end{array} \right) \chi_l.
\end{equation}

\section{Numerical solution of the Schr\"odinger equation} \label{sec:num_sol}
We wish to solve eq.~\eqref{schrodinger_eq} imposing regularity of $\chi$ at the origin as the boundary condition. We drop the $l$ subscript in the following. We follow and slightly modify the procedure of ref.~\cite{Ershov:2011zz} to solve the more general problem where there are $N$ scattering channels of which our problem is the $N=2$ special case. First, we write the wave function in component form as
\begin{equation} \label{alphabeta}
\chi_i = \alpha_i(x) f(p_i x) - \beta_i(x) g(p_i x)\,,
\end{equation}
where $i$ denotes the wave function component ($i=1,\,2$ in our case), $p_i$ is the momentum of channel $i$, and the functions $f(p_i x)$ and $g(p_i x)$ are solutions to the simpler equation
\begin{equation}
y'' = \left( \frac{l(l+1)}{x^2} - p_i^2 \right) y\,.
\end{equation}
We normalise the functions according to $f(x)g'(x) - f'(x)g(x) = -1$ and assume that $f(x)$ is regular and $g(x)$ is irregular at the origin. The two added degrees of freedom in eq.~\eqref{alphabeta} are removed by the constraint
\begin{equation}
f(p_i x) \alpha'_i(x) - g(p_i x) \beta'_i(x) = 0\,.
\end{equation}
For the functions $f$ and $g$, we choose
\begin{equation}
f(x) = x j_l(x),\qquad g(x) = i x h^{1}_l(x)\,,
\end{equation}
where $j_l$ is the spherical Bessel function and $h_l^1(x)$ is the spherical Hankel function, both of the first kind. We need $N$ linearly independent solutions to the differential equation and thus we promote the wave function $\chi_i$ in eq.~\eqref{alphabeta} to an $N \times N$ matrix, which can be written in components as
\begin{equation} \label{matrixeq}
\chi_{in} = f(p_i x) \alpha_{in}(x) - g(p_i x) \beta_{in}(x)\,,
\end{equation}
where $n=1, \dots ,N$ denotes the different solutions. That is, the columns in the matrix $\boldsymbol{\chi}(x)$ give independent wave functions that solve the differential equation. Since $g(x)$ is irregular at the origin, we require that $\beta_{in}(x \rightarrow 0) = 0$, while we take $\alpha_{in}(x \rightarrow 0) = \delta_{in}$ as initial condition.

Next, we introduce the matrix
\begin{equation} \label{Mmat}
\boldsymbol{M}(x) = \boldsymbol{\beta}(x)\boldsymbol{\alpha}^{-1}(x)\,,
\end{equation} 
which allows us to write eq.~\eqref{matrixeq} as
\begin{equation}
\boldsymbol{\chi}= \left[ \boldsymbol{f}(p_i x) - \boldsymbol{g}(p_i x) \boldsymbol{M}(x) \right] \boldsymbol{\alpha}(x)\,,
\end{equation}
where $\boldsymbol{f}$ and $\boldsymbol{g}$ are diagonal matrices with entries $f(p_i x)$ and $g(p_i x)$ in row $i$, respectively. Multiplying the solution by $\boldsymbol{\alpha}^{-1}(x)$ from the right gives a new wave function $\boldsymbol{\xi}$,
\begin{equation} \label{xi_sol}
\boldsymbol{\xi} = \boldsymbol{f}(p_i x) - \boldsymbol{g}(p_i x) \boldsymbol{M}(x)\,.
\end{equation}
The usefulness of defining $\boldsymbol{\xi}$ becomes apparent in the limit of large $x$, where the asymptotic limits of the functions $f$ and $g$ has kicked in,
\begin{equation}
\lim_{x\to\infty} f(x) = \frac{ (-i)^{l+1} e^{ix} + (i)^{l+1} e^{-ix} }{2} \, , \qquad \lim_{x\to\infty} g(x) = (-i)^{l+2} e^{ix} \, .
\end{equation}
The relation between $R_l$ and $\chi$ in eq.~\eqref{R_chi_link} gives $\boldsymbol{\xi}$ a factor $1/x$ when resubstituting. When this factor multiplies eq.~\eqref{xi_sol}, a comparison in the large $x$ limit to the expression inside the brackets of eq.~\eqref{wave_funcb} makes it clear that the $i'th$ column of $\boldsymbol{\xi}$ represents a scattered wave-function with the $i$'th state being the incoming one. In our 2-state case, the first column represents an incoming $\chi \chi$ state being scattered while the second column represents $\chi^* \chi^*$ scattering. However, the normalisation is not correct nor do each wave-function share the same normalisation factor. Multiplying the wave-function in column $i$ by $-((-i)^{l+2} p_i )^{-1}$ brings it to exactly the form of eq.~\eqref{wave_funcb}. Following this reasoning leads us to the conclusion that our scattering amplitudes $f_{X,l}$ and $f_{Y,l}$ in the case of $\chi \chi$ scattering are
\begin{equation} \label{amps_chichi}
f_{X,l} = M_{11}(x_{\infty})/a, \qquad f_{Y,l} = M_{21}(x_{\infty})/a \, ,
\end{equation}
where $x_\infty$ is large. If the up-scattering channel is open ($p_i^2 > 0$), the second column tells us that the amplitudes for elastic $\chi^* \chi^*$ scattering and down-scattering are given by
\begin{equation} \label{amps_chi*chi*}
f_{X,l} = M_{12}(x_{\infty})/c, \qquad f_{Y,l} = M_{22}(x_{\infty})/c \, .
\end{equation}
The last step is to introduce a matrix $\boldsymbol{U}$ defined as
\begin{equation} \label{M_U_coupling}
\boldsymbol{U} = \boldsymbol{f}\boldsymbol{g} - \boldsymbol{g}\boldsymbol{M}\boldsymbol{g}\, 
\end{equation}
which allows one to write the wave function in terms of $\boldsymbol{U}$ as
\begin{equation}
\chi_{in} = U_{ij} \frac{\alpha_{jn}}{g(p_j x)} \, .
\end{equation}
Using these substitutions, one can show that the original Schr\"odinger equation~\eqref{schrodinger_eq} reduces to two first order differential equations, one of which is (see ref.~\cite{Ershov:2011zz}):
\begin{equation} \label{diff_eq_U}
U'_{ij} = p_i \delta_{ij} + p_i \left[ \frac{g'_i}{g_i} + \frac{g'_j}{g_j} \right] U_{ij} - U_{il} \frac{\hat{V}_{lm}}{p_l} U_{mj}\, , \,\,\quad U_{ij}(x_0) = f(p_i x_0)g(p_i x_0) \delta_{ij} \, .
\end{equation}
where $g'_i = dg(p_i x)/d(p_i x)$, $U_{ij}(x_0)$ is the initial condition at a very small $x_0$ and $\hat{V}$ is the potential of eq.~\eqref{V_pot} without mass splittings,
\begin{equation}
\hat{V}(x) = \left(
  \begin{array}{ c c }
     0 & -\frac{1}{x} e^{-x/b} \\
     -\frac{1}{x} e^{-x/b} & 0
  \end{array} \right)\,.
\end{equation}
The second set of differential equations allows one to determine $\boldsymbol{\alpha}$, which along with $\boldsymbol{M}$ in eq.~\eqref{Mmat} allows one to compute $\boldsymbol{\beta}$. These are necessary to determine the Sommerfeld enhancement factors but not to compute the self-interacting cross sections. We will solve the differential equation eq.~\eqref{diff_eq_U} for $\boldsymbol{U}$, find $\boldsymbol{M}$ from eq.~\eqref{M_U_coupling}, and compute the amplitudes as given in eqs.~\eqref{amps_chichi}~and~\eqref{amps_chi*chi*}. In this formalism, the amplitudes for the scattering in the excited state will be given in terms of the velocity of the lower mass state. Substituting $\delta \rightarrow -\delta$ trades places between $a$ and $c$ throughout the calculation to give the scattering cross sections in terms of the incoming $\chi^*$ velocity instead.

\end{document}